\RequirePackage[l2tabu,orthodox]{nag}
\documentclass[envcountsect,envcountsame]{llncs}
\pdfoutput=1

\usepackage[T1]{fontenc}
\usepackage{fixltx2e}
\usepackage{microtype} 
\usepackage{xspace}
\usepackage{multicol}
\usepackage{eurosym}

\makeatletter
\newcommand\etc{etc\@ifnextchar.{}{.\@}\xspace}
\newcommand\ie{i.e.\@\xspace}  
\newcommand\eg{e.g.\@\xspace}
\newcommand\cf{cf.\@\xspace}
\makeatother

\usepackage{bbm}
\usepackage{graphicx}

\newcommand{\inlinegraphic}[2]{
  \dimendef\grafheight=255\dimendef\grafvshift=254
  \grafheight=#1
  \grafvshift=-0.5\grafheight
  \advance\grafvshift by 0.5ex
  \raisebox{\grafvshift}{\includegraphics[height=\grafheight]{images/#2}\xspace}
}

\newcommand{\ninlinegraphic}[2][1.0]{
  \dimendef\grafheight=255\dimendef\grafvshift=254
  \setbox0 = \hbox{\scalebox{#1}{\includegraphics{images/#2}}}
  \grafheight=\the\ht0
  \grafvshift=-0.5\grafheight
  \advance\grafvshift by 0.5ex
  \raisebox{\grafvshift}{\includegraphics[height=\grafheight]{images/#2}\xspace}
}

\usepackage{latexsym}
\usepackage{amssymb}
\usepackage{amsmath} 
\usepackage{stmaryrd}
\usepackage{mathtools}

\newcommand{\iso}{\cong}
\newcommand{\isomorphism}{\cong}

\newcommand{\sizeof}[1]{
  \left|#1\right|}

\newcommand{\bra}[1]{
    \ensuremath{\left\langle #1 \right|}\xspace}
\newcommand{\ket}[1]{
    \ensuremath{\left|  #1 \right\rangle}\xspace}

\newcommand{\zxcalculus}{\textsc{zx}-calculus\xspace}

\usepackage{diagrams} 
\newarrow{Equals}{}{=}{}{=}{}

\ifx\numericids\undefined
\newcommand{\id}[1]{\ensuremath{\mathrm{id}_{#1}}}
\else
\newcommand{\id}[1]{\ensuremath{1_{#1}}}
\fi

\newcommand{\catC}{\ensuremath{\mathcal{C}}\xspace}
\newcommand{\catD}{\ensuremath{\mathcal{D}}\xspace}

\newcommand{\op}{\ensuremath{^{\mathrm{op}}}\xspace}

\newcommand{\CC}{\catC}

\newcommand{\catname}[1]{\ensuremath{\mathbf{#1}}\xspace}

\newcommand{\fset}{\catname{FinSet}}

\newcommand{\finset}{\fset}

\newcommand{\fdhilb}{\catname{fdHilb}} 
\newcommand{\fdHilb}{\fdhilb}

\newcommand{\Ab}{\catname{Ab}}
\newcommand{\Grp}{\catname{Grp}}
\newcommand{\PRO}{\catname{PRO}}
\newcommand{\PROP}{\catname{PROP}}
\newcommand{\dPRO}{\ensuremath{\text{\dg-\PRO}}\xspace}
\newcommand{\dPROP}{\ensuremath{\text{\dg-\PROP}}\xspace}

\newcommand{\ssspan}[1]{\ensuremath{\catname{Span}(#1)}\xspace}
\newcommand{\cospan}[1]{\ensuremath{\catname{Cospan}(#1)}\xspace}

\newcommand{\dg}{\ensuremath{\dag}}

\newcommand{\PP}{\ensuremath{\textbf{P}}\xspace}
\renewcommand{\CC}{\ensuremath{\textbf{C}}\xspace}
\newcommand{\MM}{\ensuremath{\textbf{M}}\xspace}

\newcommand{\TT}{\ensuremath{\textbf{T}}\xspace}
\newcommand{\FA}{\ensuremath{\textbf{F}}\xspace}
\newcommand{\BA}{\ensuremath{\textbf{B}}\xspace}
\newcommand{\HA}{\ensuremath{\textbf{HA}}\xspace}
\newcommand{\IFA}{\ensuremath{\textbf{IF}}\xspace}

\newcommand{\IFK}{\ensuremath{\textbf{IFK}}\xspace}
\newcommand{\IFKd}{\ensuremath{\textbf{IFK}_{{\mathbf d}}}\xspace}

\usepackage{hyperref}

\usepackage[undirected]{tikzquanto}

\begin{document}

\title{Interacting Frobenius Algebras are Hopf}
\author{Ross Duncan \and Kevin Dunne}
\institute{University of Strathclyde, 26 Richmond Street, Glasgow G1
1XH, UK.}




\maketitle
%


\begin{abstract}
  Theories featuring the interaction between a Frobenius algebra and a
  Hopf algebra have recently appeared in several areas in computer
  science: concurrent programming, control theory, and quantum
  computing, among others.  Bonchi, Sobocinski, and Zanasi
  \cite{Bonchi2014a} have shown that, given a suitable distributive
  law, a pair of Hopf algebras forms two frobenius algebras.  Here we
  take the opposite approach, and show that interacting Frobenius
  algebras form Hopf algebras.  We generalise \cite{Bonchi2014a} by
  including non-trivial dynamics of the underlying object---the
  so-called phase group---and investigate the effects of finite
  dimensionality of the underlying model.  We recover the system of
  Bonchi et al as a subtheory in the prime power dimensional case, 
  but the more general theory does not arise from a distributive law.
\end{abstract}
\section{Introduction}
\label{sec:introduction}

Frobenius algebras and bialgebras are structures which combine a
monoid and a comonoid on a single underlying object.  They have
a long history\footnote{See Fauser \cite{Bertfried:2013aa} for much
  detail on Frobenius algebras, including their history; for the
  history of Hopf algebras see \cite{andruskiewitch:2009:aa}.}  in
group theory, but have applications in many other areas:
natural language processing
\cite{sadrzadeh2013frobenius,sadrzadeh2014frobenius}, topological
quantum field theory \cite{Kock:TQFT:2003}, game semantics
\cite{5230577}, automata theory \cite{Worthington:2009:aa}, and
distributed computing \cite{Benson:1988:aa}, to name but a few.

In quantum computation, the bialgebraic interplay between two Frobenius algebras describes 
the behaviour of complementary observables
\cite{Coecke:2008nx,Coecke:2009aa}, a central concept in quantum
theory.  This interaction is the basis of the \zxcalculus, a formal
language for quantum computation.  Using these ideas, a significant
fraction of finite dimensional quantum theory can be developed without
reference to Hilbert spaces 
\cite{1367-2630-16-9-093021,Backens:2014aa,Backens:2015aa,backens:2015:aa,Coecke:2013lr,CDKW-lics:2012qy,Coecke:2015aa,Coecke:2013-qpl2012,Witt:2014aa,Duncan:2010aa,Duncan:2012uq,Duncan:2013lr,Duncan:2009ph,Gogioso:2015ab,Gogioso:2015aa,Gogioso:2015ac,Heunen:2013aa,Perdrix:2015aa,Ranchin:2014ab,Ranchin:2014aa,Wang:2014aa}.
Surprisingly, almost exactly the
same axioms have also appeared in totally different settings: Petri
nets \cite{Bruni:2011:captni,sobocinksi:2013:aa} and control theory
\cite{Baez2014a,Bonchi2014b}.  This combination of structures seems to
have broad relevance in computer science.

The approach of the current paper is directly inspired by the recent
work of Bonchi, Soboci\'nski, and Zanasi \cite{Bonchi2014a}, who
investigated the theory of interacting Hopf algebras\footnote{A Hopf
  algebra is a bialgebra with some extra structure; see later
  \ref{def:hopf-alg}.} and showed that Hopf algebras which obey a
certain distributive law form Frobenius algebras
\cite{Bonchi2014,Bonchi2014a}.  Using Lack's technique of composing
PROPs \cite{Lack:2004sf}, they show the resulting theory $\mathbb{IH}_R$
is isomorphic to that of linear relations\footnote{Baez and Erbele
  \cite{Baez2014a} prove the same result with different techniques.}.

Do interacting quantum observables \cite{Coecke:2009aa} admit such a
beautiful description?  In this paper we present a rational
reconstruction of theory of strongly complementary observables and
show that, except under quite restrictive circumstances, the theory does
not arise by composing PROPs via a distributive law.  Along the way
we also clarify the structure of the theory of complementary
observables and show that some assumptions used in earlier work are
unnecessary.

In the quantum context, the key insight is that an observable of some
quantum system corresponds to a Frobenius algebra on its state space
\cite{PavlovicD:MSCS08}.  Further, the state spaces have non-trivial
endomorphims giving their internal dynamics; among these there is a
\emph{phase group} for each observable, which leaves the observable
unchanged.  Since observables are fundamental to quantum theory, we
take Frobenius algebras and their phase groups as the starting point,
and freely construct $\FA G$, the PROP of a Frobenius algebra
with a given group of phases $G$.

The general plan of the paper is to begin with a pair of such
Frobenius algebras and formalise interactions
between them by imposing stronger and stronger axioms upon them.  We
produce a series of PROPs
\[
\FA \_ + \FA \_ \rOnto \IFA \rOnto \IFK
\rOnto \IFKd
\]
each more closely resembling quantum theory than its predecessor.  The
first is simply the disjoint union of two \emph{non}-interacting
observables.  The second requires that the observables be strongly
complementary; this means their corresponding Frobenius algebras
jointly form a Hopf algebra \cite{Coecke:2008nx,Coecke:2009aa}.  The
additional structure allows us to construct a ring of endomorphisms of
the generator, distinct from the phase groups, and a large class of
multiparty unitary operations, being the abstract counterpart of
quantum circuits.  The next two PROPs introduce eigenstates for the
observables, and the effect of finite dimensionality of the state
space respectively.  In the last of these, $\IFKd$, if the dimension
is a prime power the we recover the system $\mathbb{IH}_R$ of Bonchi et
al \cite{Bonchi2014a} as a subcategory.

Each of these theories is actually a functor from a suitable category
of groups, so we can freely construct a quantum-like theory with any
given dynamics.

Our motivation for studying these generalisations is to better
understand categorical quantum theory \cite{Abramsky:2008kx},
particularly with a view to the \zxcalculus.  We explicate the
necessary features of higher dimensional versions of the calculus, and
separate the algebraic foundation from model-specific details.
This will help clarify questions of completeness
\cite{1367-2630-16-9-093021,backens:2015:aa,EPTCS171.5,Perdrix:2015aa}
and also aid in the formalisation of error correcting codes
\cite{Duncan:2013lr}.  However given the interest in these structures
in other areas, we expect that a richer theory will lead to unexpected
applications elsewhere.  As a side-effect we learn that the theory
of Petri-nets differs from quantum mechanics.

Due to restriction on space the proofs are mostly omitted; see
Appendix \ref{sec:proofs}.

\section{Background}
\label{sec:background}

We assume that the reader is familiar with the theory of monoidal
categories; all monoidal structures here are taken to be strict.  We
will employ diagrammatic notation throughout the paper; see Selinger
\cite{Selinger:2009aa}.  Our convention is to read diagrams from top
to bottom however, since we operate in a \dg-category, every diagram
can be read equally well from bottom to top; the reader who chooses to
do so will need to add the involutive prefix ``co'' throughout the
text themselves.

\begin{definition}  \label{def:dagger}
  A \emph{\dg-category} is a category \catC equipped with a functor
  $\dag: \catC\op \to \catC$ which is involutive and acts as the
  identity on objects.
\end{definition}

A morphism $f:A\to B$ in a \dg-category is called \emph{unitary} if
$f^\dag : B \to A$ is the two-sided inverse of $f$;  it is a
\emph{unitary embedding} if $f^\dag$ is a left-inverse only; it is
\emph{self-adjoint} if $f = f^\dag$.

\begin{remark}
  A groupoid is a \dg-category in which every morphism is unitary; in
  particular every group can be viewed as a one-object \dg-category.
\end{remark}

A functor $F:\catC \to \catD$ between \dg-categories is a
\dg-functor if $(Ff)^\dag = F(f^\dag)$ for all arrows $f$.  A
(symmetric) monoidal \dg-category is called \dg-(symmetric)
monoidal if $- \otimes - : \catC \times \catC \to \catC$ is a
\dg-functor, and all the canonical isomorphisms of the monoidal
structure are unitary.


The main example of interest is \fdhilb, the category of finite
dimensional Hilbert spaces over ${\mathbb C}$ and linear maps; given
$f:A\to B$, $f^\dg:B\to A$ is the usual Hermitian adjoint.

We now turn our attention to PROPs. This material largely follows
\cite{Lack:2004sf,Bonchi2014a}.

\begin{definition} \label{def:PRO} \label{def:PROP} A \emph{product 
category}, abbreviated \emph{PRO}, is a
  strict monoidal category whose objects are generated by a single
  object under the tensor product; or equivalently, whose objects are
  the natural numbers.  A \emph{product and permutation category}, abbreviated 
\emph{PROP}, is a symmetric PRO.  A
  \emph{\dg-PRO} or \emph{\dg-PROP} is a PRO (respectively PROP)
  which is also a \dg-monoidal category.
\end{definition}

Given any strict monoidal category \catC the full subcategory generated by a
single object under tensor is a PRO.  In particular, for any natural
number $D$ we can consider the full subcategory of \fdhilb generated
by $\mathbb{C}^D$ under the tensor product.  For $D=2$ this gives the
usual setting of quantum computing.

For a PRO \TT and a strict monoidal category \catC, a
\TT-\emph{algebra} in \catC is a strict monoidal functor from \TT
to \catC.  We will abuse notation and refer to the algebra by the name
of its generating object in \catC.  A morphism between PROs is an
algebra which is the identity on objects.  Therefore we have a
category \catname{PRO} of PROs and their morphisms.  The same can be
done for PROPs, \dg-PROs, and \dg-PROPs by requiring that the functor
is symmetric monoidal and/or dagger as appropriate.


Let \PP be the PRO whose morphisms $n\to n$ are the permutations on
$n$ elements, with no morphisms $n \to m$ if $n \neq m$.  \PP is
groupoid, hence also a \dg-category.  We can understand the
category \dPROP as the coslice category $\PP/\text{\dg-\PRO}$.  The
coproduct $\TT_1 + \TT_2$ in \dPROP is given by
the pushout of $\TT_1 \leftarrow \PP \to \TT_2$ in \dPRO since the
symmetric structure has to agree in both.

In this paper we are concerned with PROPs which are presented
syntactically.  The arrows of the PROP will be constructed by
composition and tensor from the elements of a monoidal signature
$\Sigma$ and a set $E$ of equations between terms of the same type.
Equality is then the least congruence generated by $E$ and
the equations of the symmetric monoidal structure.  From this point of
view the coproduct $\TT_1+\TT_2$ is given by the pair
$(\Sigma_1+\Sigma_2, E_1 + E_2)$.

The coproduct is not an especially exciting operation: we need to
combine PROPs \emph{and} make them interact. Lack's method of
composing PROPs via distributive laws is a particularly elegant approach
\cite{Lack:2004sf}.  We will skip the details here, but given two
PROPs $\TT_1$ and $\TT_2$, a distributive law $\lambda : \TT_2;\TT_1
\to \TT_1;\TT_2$ is a set of directed equations $(f_2,f_1)\to
(f_1',f_2')$ commuting morphisms of $\TT_2$ past those of $\TT_1$.
The composite PROP $\TT_1;\TT_2$ has morphisms of the form $n
\rTo^{f_1} z \rTo^{f_2} m$ where $f_1$ is an arrow of $\TT_1$ and
$f_2$ of $\TT_2$; its syntactic presentation is that of $\TT_1 +
\TT_2$ with the additional equations of $\lambda$.

\begin{example}\label{ex:prop-PG}
  As a simple example we can view \PP as a PRO with a single generator
  $c : 2 \to 2$ quotiented by $c^2 = \id{}$ and the usual hexagon
  diagrams.  Let $G$ be a group; we define $G^{\times}$ to be the PRO
  with hom-sets $G^\times(n,n) = \prod_n G$, and $G^\times(n,m) =
  \emptyset$ if $n \neq m$.  Composition is done component-wise in
  $G$.  The generators of $G^\times$ are just the elements $g : 1 \to
  1$ for each $g \in G$ quotiented by the equations of $G$.  We can
  define the composite $\PP;G^\times$ via the distributive law:
  \[
  \lambda : 
\beginpgfgraphicnamed{natural-twist1}
\InputIfFileExists{natural-twist1.tikz}{}{\input{./figures/natural-twist1.tikz}}
\endpgfgraphicnamed \;=\;\;
\beginpgfgraphicnamed{natural-twist2}
\InputIfFileExists{natural-twist2.tikz}{}{\input{./figures/natural-twist2.tikz}}
\endpgfgraphicnamed 
  \]
  for each $g_1$ and $g_2$ in $G$.  This yields the PRO -- actually a
  \dg-PROP -- whose morphisms $n\to n$ are a permutation on $n$
  followed by an $n$-vector of elements of $G$.  It's easy to see that
  this construction yields a functor $\PP : \Grp \to \dPROP$.  Notice
  that $\PP G$ is again a groupoid, and every morphism is unitary.
\end{example}

\begin{example}\label{ex:prop-bialgs-and-Falgs}
  A second cluster of examples, stolen shamelessly from
  \cite{Lack:2004sf}, provides the main structures of interest of this
  paper.  Let \MM denote the PROP of \emph{commutative monoids}; it
  has two generators, $\mu:2\to1$ and $\eta:0:\to1$, which we write
  graphically as \gmult and \gunit, subject to the equations:

\begin{equation}\label{eq:monoid}\tag{M}
\beginpgfgraphicnamed{assoc1}
\InputIfFileExists{assoc1.tikz}{}{\input{./figures/assoc1.tikz}}
\endpgfgraphicnamed
  = 
\beginpgfgraphicnamed{assoc2}
\InputIfFileExists{assoc2.tikz}{}{\input{./figures/assoc2.tikz}}
\endpgfgraphicnamed
  \qquad
\beginpgfgraphicnamed{unit2}
\InputIfFileExists{unit2.tikz}{}{\input{./figures/unit2.tikz}}
\endpgfgraphicnamed = \;%
\beginpgfgraphicnamed{id}
\begin{tikzpicture}
	\begin{pgfonlayer}{nodelayer}
		\node [style=none] (0) at (0, 0.5) {};
		\node [style=none] (1) at (0, -0.5) {};
	\end{pgfonlayer}
	\begin{pgfonlayer}{edgelayer}
		\draw (0.center) to (1.center);
	\end{pgfonlayer}
\end{tikzpicture}}
\endpgfgraphicnamed\;
  \qquad
\beginpgfgraphicnamed{comm1}
\InputIfFileExists{comm1.tikz}{}{\input{./figures/comm1.tikz}}
\endpgfgraphicnamed = %
\beginpgfgraphicnamed{comm2}
\InputIfFileExists{comm2.tikz}{}{\input{./figures/comm2.tikz}}
\endpgfgraphicnamed
\end{equation}
We can define the PROP of cocommutative comonoids as $\CC = \MM\op$.
The generators are $\delta: 1 \to 2$ and $\epsilon: 1\to0$; the
equations are those of (\ref{eq:monoid}) but flipped upside down. We call these 
equations C.
Bialgebras and Frobenius algebras combine a monoid and comonoid in
different ways; both can be built using distributive laws
between \MM and \CC.
\end{example}

\begin{example}
The PROP \BA of \emph{commutative bialgebras} is constructed via a
distributive law $\lambda_B : \MM;\CC \to \CC;\MM$ generated by the
equations
\begin{align*}\label{eq:bialgebra}\tag{B}
\beginpgfgraphicnamed{bialg2lhs-mono}
\InputIfFileExists{bialg2lhs-mono.tikz}{}{\input{./figures/bialg2lhs-mono.tikz}}
\endpgfgraphicnamed = \;
\beginpgfgraphicnamed{bialg2rhs-mono}
\InputIfFileExists{bialg2rhs-mono.tikz}{}{\input{./figures/bialg2rhs-mono.tikz}}
\endpgfgraphicnamed 
\qquad\quad
\beginpgfgraphicnamed{bialg-copy-lhs-mono}
\begin{tikzpicture}
	\begin{pgfonlayer}{nodelayer}
		\node [style=green vertex] (0) at (0, 0) {};
		\node [style=green vertex] (1) at (0, 0.5) {};
		\node [style=none] (2) at (-0.25, -0.5) {};
		\node [style=none] (3) at (0.25, -0.5) {};
	\end{pgfonlayer}
	\begin{pgfonlayer}{edgelayer}
		\draw (1) to (0);
		\draw (0) to (2.center);
		\draw (0) to (3.center);
	\end{pgfonlayer}
\end{tikzpicture}}
\endpgfgraphicnamed &= \;
\beginpgfgraphicnamed{bialg-copy-rhs-mono}
\begin{tikzpicture}
	\begin{pgfonlayer}{nodelayer}
		\node [style=green vertex] (0) at (-0.25, 0.25) {};
		\node [style=green vertex] (1) at (0.25, 0.25) {};
		\node [style=none] (2) at (-0.25, -0.25) {};
		\node [style=none] (3) at (0.25, -0.25) {};
	\end{pgfonlayer}
	\begin{pgfonlayer}{edgelayer}
		\draw (0) to (2.center);
		\draw (3.center) to (1);
	\end{pgfonlayer}
\end{tikzpicture}}
\endpgfgraphicnamed 
\qquad\quad
\beginpgfgraphicnamed{bialg-cocopy-lhs-mono}
\begin{tikzpicture}
	\begin{pgfonlayer}{nodelayer}
		\node [style=green vertex] (0) at (0, 0) {};
		\node [style=green vertex] (1) at (0, -0.5) {};
		\node [style=none] (2) at (-0.25, 0.5) {};
		\node [style=none] (3) at (0.25, 0.5) {};
	\end{pgfonlayer}
	\begin{pgfonlayer}{edgelayer}
		\draw (1) to (0);
		\draw (0) to (2.center);
		\draw (0) to (3.center);
	\end{pgfonlayer}
\end{tikzpicture}}
\endpgfgraphicnamed = \;
\beginpgfgraphicnamed{bialg-cocopy-rhs-mono}
\begin{tikzpicture}
	\begin{pgfonlayer}{nodelayer}
		\node [style=green vertex] (0) at (-0.25, -0.25) {};
		\node [style=green vertex] (1) at (0.25, -0.25) {};
		\node [style=none] (2) at (-0.25, 0.25) {};
		\node [style=none] (3) at (0.25, 0.25) {};
	\end{pgfonlayer}
	\begin{pgfonlayer}{edgelayer}
		\draw (0) to (2.center);
		\draw (3.center) to (1);
	\end{pgfonlayer}
\end{tikzpicture}}
\endpgfgraphicnamed \\
\beginpgfgraphicnamed{green-bone}
\begin{tikzpicture}
	\begin{pgfonlayer}{nodelayer}
		\node [style=green vertex] (0) at (0, -0.25) {};
		\node [style=green vertex] (1) at (0, 0.25) {};
	\end{pgfonlayer}
	\begin{pgfonlayer}{edgelayer}
		\draw (1) to (0);
	\end{pgfonlayer}
\end{tikzpicture}}
\endpgfgraphicnamed &= \;
\beginpgfgraphicnamed{dashedbox}
\InputIfFileExists{dashedbox.tikz}{}{\input{./figures/dashedbox.tikz}}
\endpgfgraphicnamed 
\end{align*}
where the dashed box represents the empty diagram.
\end{example}

\begin{example}\label{exmp:FrobeniusAlgebraPROP}
The PROP \FA of \emph{Frobenius algebras} is also defined by 
 distributive law, 
$\lambda_F:\CC;\MM \to \MM;\CC$, given by the equations:
\begin{equation}\label{eq:frobenius}\tag{F}
\beginpgfgraphicnamed{Frob1}
\InputIfFileExists{Frob1.tikz}{}{\input{./figures/Frob1.tikz}}
\endpgfgraphicnamed \; = %
\beginpgfgraphicnamed{Frob2}
\InputIfFileExists{Frob2.tikz}{}{\input{./figures/Frob2.tikz}}
\endpgfgraphicnamed = \; %
\beginpgfgraphicnamed{Frob3}
\InputIfFileExists{Frob3.tikz}{}{\input{./figures/Frob3.tikz}}
\endpgfgraphicnamed
  \qquad\qquad
\beginpgfgraphicnamed{special}
\InputIfFileExists{special.tikz}{}{\input{./figures/special.tikz}}
\endpgfgraphicnamed \; = \;\;%
\beginpgfgraphicnamed{id}
\begin{tikzpicture}
	\begin{pgfonlayer}{nodelayer}
		\node [style=none] (0) at (0, 0.5) {};
		\node [style=none] (1) at (0, -0.5) {};
	\end{pgfonlayer}
	\begin{pgfonlayer}{edgelayer}
		\draw (0.center) to (1.center);
	\end{pgfonlayer}
\end{tikzpicture}}
\endpgfgraphicnamed
\end{equation}
This is not the most general form of Frobenius algebra. More
accurately, \FA is the PROP of \emph{special commutative Frobenius
  algebras}; the last equation above is what makes them ``special''.
Throughout this paper the reader should understand the term
``Frobenius algebra'' to mean ``special commututive \dg-Frobenius
algebra'', usually abbreviated \dg-SCFA.  Rosebrugh, Sabadini, and
Walters call the same structure a \emph{separable commutative algebra}
\cite{Rosebrugh:2004aa}.

\end{example}

By defining the PROP $\FA$ in Example \ref{exmp:FrobeniusAlgebraPROP}
via the distributive law $\lambda_F$ we can see the following ``Spider
Theorem'' \cite{Coecke2006POVMs-and-Naima}, which establishes a normal
form for morphisms in the PROP $\FA$. In particular every morphism in
$\FA$ can be expressed as the composition of a morphism in $\MM$
followed by a term in $\CC$

\begin{theorem}[Spider Theorem]
  \label{thm:spider}
  Let $f:m\to n$ be a morphism in \emph{\FA}; if the
  graphical form of $f$ is connected then $f = \delta_n \circ \mu_m$ where 
  \[
  \delta_0 := \epsilon \qquad\qquad \delta_{k+1} := (\delta_k \otimes
  \id{A}) \circ \delta
  \]
  and $\mu_m$ is defined dually.
\end{theorem}
With this in mind we define a ``spider'' $\gdot^{m}_{n} :=
\delta_n\circ\mu_m$ as a tree of $m$ multiplies followed by a co-tree
of $n$ comultiplies.  We can view \FA as the category of spiders,
where composition means fusing connected spiders and removing any
self-loops.

\begin{remark}\label{rem:cospans}
  We note that all of these PROPs also have ``semantic''
  presentations: \MM to equivalent to \finset, the skeletal category
  of finite sets and functions, while \BA and \FA are equivalent to
  \ssspan\finset and \cospan\finset respectively.  The spider theorem
  is equivalent to this last fact.  See \cite{Rosebrugh:2004aa} and
  \cite{Lack:2004sf}
\end{remark}



For any category $\TT$, one can view $\TT+\TT\op$ as a \dg-category, hence
in all of the above we may assume that $\delta = \mu^\dg$ and
$\epsilon = \eta^\dg$.  However it is not always desirable to do so.
Whether we view $\CC;\MM$ as a PROP or a \dg-PROP makes a difference
when considering it algebras in some other \dg-category.
In the sequel we will take \FA as the \dg-PROP of \dg-Frobenius
algebras but will ignore the \dg-structure of \BA.

\section{The Standard Model}
\label{sec:standard-model}

The combination of Frobenius and Hopf algebras arises naturally in the
study of quantum observables.  In this section we present a class of
concrete examples that exist in every finite dimensional complex
Hilbert space.  The starting point is this theorem of Coecke,
Pavlovic, and Vicary \cite{PavlovicD:MSCS08}:

\begin{theorem}\label{thm:dscfa-are-ONbases}
  In \fdHilb, $(\gdelta, \gepsilon)$ is a \dg-SCFA on $A$ iff  
  \[
  \gdelta: \ket{a_i} \mapsto \ket{a_i} \otimes \ket{a_i} 
  \qquad\qquad 
  \gepsilon : \ket{a_i}  \mapsto 1.
  \]
  for some orthonormal basis $\{\ket{a_i}\}_i$ of $A$.
\end{theorem}

For any coalgebra the elements copied by $\delta$ -- the $\ket{a_i}$
in the theorem above -- are called \emph{set-like}.  So given an
orthonormal basis $\ket{0},\ldots, \ket{D-1}$ for the Hilbert space
$\mathbb{C}^D$ we get a $\dg$-SCFA defined as above, whose set-like
elements are $\ket{n}$.  We can construct another Frobenius algebra by
viewing this basis as the elements of the additive group $\mathbb{Z}_D$
and forming the group algebra:
\[
\rmu : \ket n \otimes \ket m \mapsto \ket{n+m} 
\qquad\qquad
\reta : \ket 0 \mapsto 1
\]
This is again a \dg-Frobenius algebra, although it is
\emph{quasi-special} \cite{Gogioso:2015aa}
rather than special: we have $\mu\circ\delta = D\cdot\id{}$ rather
than the usual ``special'' equation.


This pair of Frobenius algebras are pair-wise Hopf algebras (see Def. 
\ref{def:hopf-alg}) in the sense that 
$(\rmu, \gdelta)$ is a Hopf algebra, as is $(\gmu, \rdelta)$.

\begin{remark}\label{rem:group-algebra}
 Any finite 
abelian group $G$ determines such a tuple $(\gmu,\gdelta,\rmu, \rdelta)$, see 
Table \ref{table}, which we will denote $\mathbb{C}G$ and call the 
\emph{group algebra} of $G$.    
\end{remark}
Such pairs
of quantum observables are called \emph{strongly complementary}
\cite{CDKW-lics:2012qy} and are closely related to the Fourier
transform \cite{Gogioso:2015aa}.

\begin{table*}
  \makebox[\textwidth][c]{%
    \begin{tabular}{ l | c | c |}
      & Hopf & Hopf \\ \hline 
      Frobenius & $\rmu : : \ket{n} \otimes \ket{m} \mapsto \ket{n+m}$ & 
      $\rdelta::\ket{n} \mapsto \sum\limits_{m+m'=n}\ket{m} 
      \otimes \ket{m'}$ \\ 
      & $\reta = \ket{e}$ & $\repsilon = \bra{e}$ \\ 
      \hline
      Frobenius & $\gdelta :: \ket{n} \mapsto \ket{n} \otimes \ket{n}$ & 
$\gmu:: 
      \ket{n} \otimes \ket{m}  \mapsto \ket{n}$ if $g=h$, $0$ otherwise\\ 
      & $\gepsilon =\sum\limits_{n \in G}\bra{n}$ & $\geta 
=\sum\limits_{n \in 
        G}\ket{n}$ \\ \hline
    \end{tabular}
  }
\caption{Complex Group Algebra}\label{table}
\end{table*}

Recall that the \emph{dual group} $G^{\wedge}$ of a finite abelian group $G$ is 
the set of group homomorphisms from $G$ into the circle group
of unit complex numbers, with multiplication in $G^{\wedge}$ computed
point-wise. We have $G \cong G^{\wedge}$, although this isomorphism is
not natural.
The set-like elements of $\rdelta$ are in 1-1 corresondence with elements of 
the $G^{\wedge}$, in particular, for a group character $\chi$,
\[
\ket \chi := \sum_{g \in G}\chi(g) \ket{g}
\]
is set-like for \rdelta.
%
%
%
Distinct $\ket\chi, \ket{\chi'}$ are orthogonal, so by rescaling we
obtain an orthonormal basis, and via Theorem
\ref{thm:dscfa-are-ONbases} a \dg-SCFA as required.
Moreover, in \fdHilb every pair of interacting $\dg$-SCFAs is of the form 
$\mathbb{C}G$ for a finite abelian group $G$ \cite{CDKW-lics:2012qy}.




Aside from providing some intuition for what a pair of interacting
Frobenius algebras might be, we will use these examples as a source of
counter-models to show that certain equations do not hold in the
syntactic PROPs we define in the main body of the paper.  Most
of this holds for group algebras over arbitrary fields; see Appendix
\ref{sec:vector-space-models}.

\section{Frobenius Algebras and Phases}
\label{sec:frobenius-algebras}

By Theorem \ref{thm:dscfa-are-ONbases}, every Frobenius algebra in
\fdHilb corresponds to a non-degenerate quantum observable: the
set-like elements of the coalgebra are the eigenstates of the
observable.  In this concrete setting, the maps which fix a given
observable are of great interest; we call them \emph{phases}. Before
developing this idea in the abstract setting we will recall some properties of 
\FA-algebras.

Let $A$ be an \FA-algebra in some category \catC; we let $\delta$,
$\mu$ etc stand for their images in \catC. The following proposition 
follows from the Spider Theorem.

\begin{proposition}\label{prop:FA-is-compact}
  The PROP \emph{\FA} is \dg-compact
  \cite{KelLap:comcl:1980}, with all objects self-dual.
  \begin{proof}
    Let $d = \gdot^0_2 = \gcup$ and $e = d^\dg$.  Then 
    \[
\beginpgfgraphicnamed{green-s-bend}
\InputIfFileExists{green-s-bend.tikz}{}{\input{./figures/green-s-bend.tikz}}
\endpgfgraphicnamed \; = \; %
\beginpgfgraphicnamed{green-id}
\begin{tikzpicture}
	\begin{pgfonlayer}{nodelayer}
		\node [style=green vertex] (0) at (0, 0) {};
		\node [style=none] (1) at (0, 0.5) {};
		\node [style=none] (2) at (0, -0.5) {};
	\end{pgfonlayer}
	\begin{pgfonlayer}{edgelayer}
		\draw (1.center) to (0);
		\draw (0) to (2.center);
	\end{pgfonlayer}
\end{tikzpicture}}
\endpgfgraphicnamed \; = \;
\beginpgfgraphicnamed{id}
\begin{tikzpicture}
	\begin{pgfonlayer}{nodelayer}
		\node [style=none] (0) at (0, 0.5) {};
		\node [style=none] (1) at (0, -0.5) {};
	\end{pgfonlayer}
	\begin{pgfonlayer}{edgelayer}
		\draw (0.center) to (1.center);
	\end{pgfonlayer}
\end{tikzpicture}}
\endpgfgraphicnamed
    \]
    by the spider theorem, which makes $1$ self-dual; the required cup
    and cap for the other objects can be easily constructed (although
    see \cite{Selinger:2010:aa} for the coherence conditions) to make
    all of \FA compact.  For \dg-compactness, we require
    \[
    (\gcup)^\dg = %
\beginpgfgraphicnamed{green-twist-cup}
\InputIfFileExists{green-twist-cup.tikz}{}{\input{./figures/green-twist-cup.tikz}}
\endpgfgraphicnamed
    \]
    which again follows from the spider theorem.
  \end{proof}
\end{proposition}

Obviously, compactness of \FA implies that any \FA-algebra is also
compact, in particular the inclusion of \FA into another PROP.  
Given a map $f:A\to A$, we can construct its ``\gdot-transpose'', by
conjugating with $d$ and $e$:
\[
f\gtrans = %
\beginpgfgraphicnamed{f-transpose}
\InputIfFileExists{f-transpose.tikz}{}{\input{./figures/f-transpose.tikz}}
\endpgfgraphicnamed
\]
The \gdot-transpose extends to an involutive contravariant functor on
any \FA-algebra $A$,
and since \FA is \dg-compact, the adjoint and the \gdot-transpose commute, and
hence we can define a covariant involution, the \emph{\gdot-conjugate}:
\[
f\gconj = (f^\dg)\gtrans = (f\gtrans)^\dg\;.
\]
We say that $f$ is \emph{\gdot-real} if $f = f\gconj$, or equivalently
if $f^\dg = f\gtrans$.  Evidently, the defining maps of the Frobenius
algebra are \gdot-real, as is the symmetry of the monoidal structure,
hence in \FA itself $f^\dg = f\gtrans$ for all $f$.  This is
not true for \FA-algebras in general.  

Before moving on we state a useful lemma.

\begin{lemma}\label{lem:frob-morphisms-are-iso}
  If a morphism $f$ commutes with both the monoid and comonoid parts
  of a Frobenius algebra, then it is invertible and $f^{-1} = f\gtrans$.
\end{lemma}

We are now ready to develop the abstract theory of phases.

\begin{definition}\label{def:phase}
  A \emph{pre-phase} for the \dg-SCFA $(A,\delta,\mu)$ is a map $\alpha:A\to
  A$ which acts as a strength for the multiplication:

  \begin{equation}\label{eq:phase-commute}\tag{$\mathbf{\Phi}$}
    %
\beginpgfgraphicnamed{prephase1}
\InputIfFileExists{prephase1.tikz}{}{\input{./figures/prephase1.tikz}}
\endpgfgraphicnamed \;=\;%
\beginpgfgraphicnamed{prephase2}
\InputIfFileExists{prephase2.tikz}{}{\input{./figures/prephase2.tikz}}
\endpgfgraphicnamed
\end{equation}
  A pre-phase is a \emph{phase} if it is unitary.
\end{definition}

\begin{definition}\label{def:Lambda}
  Let $\psi : I \to A$ and define $\Lambda(\psi) : A \to A$ by 
  \[
  \Lambda(\psi) : \psi \mapsto \mu \circ ( \psi \otimes \id{})
  \qquad\; 
\beginpgfgraphicnamed{point-psi}
\begin{tikzpicture}
	\begin{pgfonlayer}{nodelayer}
		\node [style=point] (0) at (0, 0.25) {$\psi$};
		\node [style=none] (1) at (0, -0.25) {};
	\end{pgfonlayer}
	\begin{pgfonlayer}{edgelayer}
		\draw (0) to (1.center);
	\end{pgfonlayer}
\end{tikzpicture}
}
\endpgfgraphicnamed \;\mapsto\;%
\beginpgfgraphicnamed{psi-phase}
\InputIfFileExists{psi-phase.tikz}{}{\input{./figures/psi-phase.tikz}}
\endpgfgraphicnamed
  \;.
  \]
\end{definition}
It follows immediately from this definition that $\Lambda(\psi)$ is a
pre-phase.  If $\Lambda(\psi)$ is in fact a phase, then we say that
$\psi$ is \emph{\gdot-unbiased}.


\begin{lemma}\label{lem:phases-and-unb-points}
  Let $\alpha:A\to A$ be a phase. Then there exists $\psi : I \to A$ such
  that 
  \begin{enumerate}
  \item $\alpha = \Lambda(\psi)$;
  \item $\alpha\gtrans = \alpha$;
  \item $\alpha^\dg = \Lambda(\psi\gconj)$;  
  \item $\mu( \psi \otimes \psi\gconj) = \eta$.
  \end{enumerate} 
\end{lemma}

\begin{corollary}\label{cor:phases-are-nice}
  If $\alpha$ is a phase, then so is $\alpha^\dg$.
\end{corollary}

\begin{lemma}\label{lem:phases-are-abelian-group}
  Let $\Phi$ denote the set of phases, and $\mathcal{U}$ denote
  the unbiased points; then 
  $(\Phi, \circ, \id{}, ()^\dg)$ and 
  $(\mathcal{U}, \mu, \eta, ()\gconj)$ are isomorphic abelian groups.
\end{lemma}

We will now consider the \dg-PROP which is generated by a \dg-SCFA
with a prescribed group of phases i.e. where $(\Phi, \circ, \id{},
()^\dg)\cong G$ for some abelian group $G$. As in example \ref{ex:prop-PG}, 
given the abelian group $G$ we can construct the PROP $\PP G$. We might then 
hope to compose
the PROPs \FA and $\PP G$ using a distributive law \cite{Lack:2004sf},
but this is impossible.  However, we can form the desired PROP via an
\emph{iterated distributive law} \cite{Cheng2011PSP8239257} (see
Appendix \ref{sec:IteratedDistLaws}).  To combine \FA and $\PP G$ we
compose the PROPs \MM, \CC and $\PP G$ pairwise via distributive laws,
and then show that these distributive laws interact nicely with one
another in such a way to yield the desired PROP.

\begin{lemma}
\begin{enumerate}
\item The PROPS \emph{\MM} and \emph{$\PP G$} can be composed via a 
distributive law $\sigma: \emph{\PP} G; \emph{\MM} \to \emph{\MM} ; \emph{\PP} 
G$, yielding a PROP presented by the equations of \emph{$\MM + \PP G$} and 
equation \emph{(P1)};

\begin{equation}\label{eqn:IteratedDistLawForFG1}\tag{P1}               
\beginpgfgraphicnamed{iterateddistlaw1}
\InputIfFileExists{iterateddistlaw1.tikz}{}{\input{./figures/iterateddistlaw1.tikz}}
\endpgfgraphicnamed=%
\beginpgfgraphicnamed{iterateddistlaw2}
\InputIfFileExists{iterateddistlaw2.tikz}{}{\input{./figures/iterateddistlaw2.tikz}}
\endpgfgraphicnamed \end{equation}\\
\item The PROPs \emph{\CC} and \emph{$\PP G$} can be composed via a 
distributive law $\rho :\emph{\CC} ; \emph{\PP} G \to \emph{\PP} G ; 
\emph{\CC}$, yielding a PROP presented by the equations of \emph{$\CC + \PP G$} 
and equation \emph{(P2)}.
\end{enumerate}

\begin{equation}\label{eqn:IteratedDistLawForFG2}\tag{P2}              
\beginpgfgraphicnamed{iterateddistlaw3}
\InputIfFileExists{iterateddistlaw3.tikz}{}{\input{./figures/iterateddistlaw3.tikz}}
\endpgfgraphicnamed=%
\beginpgfgraphicnamed{iterateddistlaw4}
\InputIfFileExists{iterateddistlaw4.tikz}{}{\input{./figures/iterateddistlaw4.tikz}}
\endpgfgraphicnamed
\end{equation}

%

\end{lemma}

Recall that the PROP \FA is defined by a distributed law $\lambda_F: \CC ; \MM 
\to \MM ; \CC$ (Example \ref{exmp:FrobeniusAlgebraPROP}).

\begin{theorem}
The distributive laws $\lambda_F$, $\rho$ and $\sigma$ form a distributive 
series of monads (Definition \ref{def:DistSeriesOfMonads}), and hence determine 
a PROP \emph{$\FA G$} presented by the 
equations of \emph{$\MM + \PP G + \CC$} and equations 
\emph{(\ref{eqn:IteratedDistLawForFG1})}, 
\emph{(\ref{eqn:IteratedDistLawForFG2})} and \emph{(\ref{eq:frobenius})}.
\begin{proof}
We need to check that the Yang-Baxter diagram of Definition 
\ref{def:DistSeriesOfMonads} commutes.
The result is then a direct application of the main theorem in 
\cite[Theorem 2.1]{Cheng2011PSP8239257}.
\end{proof}
\end{theorem}

Note that every \FA-algebra has a group of phases, although it may be
the trivial group.  We now construct the PROP of Frobenius algebras with a
\emph{given} phase group $G$.  Take any abelian group $G$ and consider
the \dg-PROP $\PP G$ as earlier; then the distributive laws $\rho$, $\sigma$ 
and $\lambda_{F}$ allow us to define the functor.
\[
\FA : \Ab \to \dPROP\;,
\]
For example $\FA 1$ is simply the original PROP of Frobenius algebras $\FA$. 
This yields an abstract 
counterpart to Theorem \ref{thm:gen-spider} as the following factorisation.

The PROPs \FA and $\PP G$ embed in $\FA G$, and equation (P1) ensures that 
the morphisms $1 \to 1$ in $\PP G$ are phases for the \dg-SCFA i.e. that they 
satisfy equation $(\mathbf{\Phi})$.

\begin{corollary}\label{thm:FG-structure}
  Let $f:n\to n'$ in \emph{$\FA G$}; then
  \[
  f = n \rTo^\nabla m \rTo^g m \rTo^\Delta n'
  \]
  where $\nabla : n \to m$ is in \emph{\MM}, $\Delta:m\to n'$ is in \emph{\CC}, 
$g :
  m\to m$ is in $G^\times$and
  $m \leq n,n'$.
\end{corollary}

Note that $\FA G$-algebras may have many more phases than those from $G$. We 
will denote the full group of phases $\Phi$, of which $G$ is necessarily a 
subgroup. Just as $\FA G$ generalises \FA, Corollary \ref{thm:FG-structure} 
lets us 
generalise the Spider Theorem.

\begin{theorem}[Generalised Spider]\label{thm:gen-spider}
  Let $f:A^{\otimes m} \to A^{\otimes n}$ be a morphism built from
  $\delta,\epsilon,\mu, \eta$, and some collection of phases $\alpha_i \in 
\Phi$ by
  composition and tensor; if the graphical form of $g$ is connected
  then $f = \delta_n \circ \alpha \circ \mu_m$ where 
  \[
  \alpha = \alpha_1 \circ \cdots \circ \alpha_k
  \]
\end{theorem}

Therefore a Frobenius algebra and its group of phases generate a
category of $\Phi$-labelled spiders.  Composition is given by fusing
connected spiders and summing their labels.

In particular, if $n = n' = 1$ in the above then $f$ is either a phase
map or a ``projector'' $\phi \circ \psi^\dg$ for a pair of unbiased
points $\phi,\psi : 0 \to 1$.  The following is a consequence of
Theorem \ref{thm:gen-spider}.

\begin{lemma}\label{lem:FG-unitaries-are-PG}
 Suppose $f : n\to n$ is unitary in \emph{$\FA G$}; then \emph{$f \in \PP 
G$}
 \end{lemma}


\section{Two Frobenius Algebras}
\label{sec:two-frob-algebr}

We briefly consider the structure of the free \dg-PROP $\FA G + \FA H$, \ie the
case of two \emph{non}-interacting Frobenius algebras.

\paragraph*{Notation}
We will adopt the convention that elements in image of the first
injection (\ie from $\FA G$) are coloured \emph{green} and the
elements in the second ($\FA H$) are coloured \emph{red}.  In
practice, the colour we call ``green'' may be light grey, and ``red''
may be dark grey depending how you read this document.

Morphisms of $\FA G + \FA H$ are alternating sequences of morphisms
from $\FA G$ and $\FA H$; \ie $f = g_1\circ h_1\circ g_2\circ h_2
\circ \cdots \circ g_n \circ h_n$ where $g_i \in \FA G$ and $h_j \in
\FA H$.  Although no equations force the two components to interact,
the spider theorem holds separately in each colour, hence any morphism
can be reduced to a 2-coloured graph, and any 2-coloured (self-loop
free) graph is valid morphism.  The following is a consequence of Lemma
\ref{lem:FG-unitaries-are-PG}.

\begin{lemma}\label{lem:unitaries-non-interacting}
  Let $u:n\to n$ be unitary in \emph{$\FA G + \FA H$}; then $u$ is in 
\emph{$\PP G
  + \PP H$}.
\end{lemma}

As a special case of the above, if $u:1\to 1$ is unitary, it is an
element of the free product of groups $G * H$.  However, unlike in
$\FA G$ this group structure is not reflected back to the points,
since we have to choose between $\gmu$ and $\rmu$ for the
multiplication, and the wrong colour merely generates the free monoid
on $G$ rather than reproducing the group structure.


In $\FA G + \FA H$ we have two distinct transposition and conjugation
operations which do not coincide, \ie $f\gtrans \neq f\rtrans$.

\begin{lemma}\label{lem:real-implies-monochrome}
  Let $f: n \to n$ be a morphism in $\FA 1 + \FA 1$; then $f$ is
  \gdot-real iff it is green and \rdot-real iff it is red.
\end{lemma}

\begin{corollary}
  \label{cor:red-and-green-implies-no-colour}
  In $\FA 1 + \FA 1$,  $f\gconj = f\rconj$ implies $f \in \PP 1$.
\end{corollary}


\section{Interacting Frobenius Algebras}
\label{sec:inter-frob-algebr}

The notion of two observables being \emph{complementary} is central to
the theory of quantum mechanics. In categorical quantum mechanics
strong complementarity is characterised by a pair of Frobenius
algebras jointly forming a Hopf algebra \cite{Coecke:2009aa}.

We now impose some equations on $\FA G + \FA H$ governing their
interaction.
We want $\FA G$ and $\FA H$ to jointly form a bialgebra
so we impose:
\begin{equation}\label{eq:bialg-laws}\tag{B}
\beginpgfgraphicnamed{bialg2lhs}
\InputIfFileExists{bialg2lhs.tikz}{}{\input{./figures/bialg2lhs.tikz}}
\endpgfgraphicnamed = \;%
\beginpgfgraphicnamed{bialg2rhs}
\InputIfFileExists{bialg2rhs.tikz}{}{\input{./figures/bialg2rhs.tikz}}
\endpgfgraphicnamed
\qquad
\beginpgfgraphicnamed{bialg-copy1}
\InputIfFileExists{bialg-copy1.tikz}{}{\input{./figures/bialg-copy1.tikz}}
\endpgfgraphicnamed = %
\beginpgfgraphicnamed{bialg-copy2}
\begin{tikzpicture}
	\begin{pgfonlayer}{nodelayer}
		\node [style=none] (0) at (0.25, -0.25) {};
		\node [style=none] (1) at (-0.25, -0.25) {};
		\node [style=red vertex] (2) at (-0.25, 0.25) {};
		\node [style=red vertex] (3) at (0.25, 0.25) {};
	\end{pgfonlayer}
	\begin{pgfonlayer}{edgelayer}
		\draw (3) to (0.center);
		\draw (2) to (1.center);
	\end{pgfonlayer}
\end{tikzpicture}}
\endpgfgraphicnamed
\qquad
\beginpgfgraphicnamed{bialg-cocopy1}
\InputIfFileExists{bialg-cocopy1.tikz}{}{\input{./figures/bialg-cocopy1.tikz}}
\endpgfgraphicnamed = %
\beginpgfgraphicnamed{bialg-cocopy2}
\begin{tikzpicture}
	\begin{pgfonlayer}{nodelayer}
		\node [style=none] (0) at (0.25, 0.25) {};
		\node [style=none] (1) at (-0.25, 0.25) {};
		\node [style=green vertex] (2) at (-0.25, -0.25) {};
		\node [style=green vertex] (3) at (0.25, -0.25) {};
	\end{pgfonlayer}
	\begin{pgfonlayer}{edgelayer}
		\draw (3) to (0.center);
		\draw (2) to (1.center);
	\end{pgfonlayer}
\end{tikzpicture}}
\endpgfgraphicnamed   
\end{equation}
We call the resulting structure a \emph{Frobenius bialgebra}:  the
pairs $(\gdelta,\gmu)$ and $(\rdelta,\rmu)$ \dg-SCFAs, while the pairs
$(\gdelta,\rmu)$ and $(\rdelta,\gmu)$ are bialgebras.

\begin{remark} \label{rem:scalars} This definition differs from the
  usual one by the presence of the scalar factor $\gepsilon\reta$ in the
  equations, and the omission of the equation:
  \begin{equation}\label{eq:scalar-is-trivial}\tag{B'}
\beginpgfgraphicnamed{rg-bone}
\begin{tikzpicture}
	\begin{pgfonlayer}{nodelayer}
		\node [style=red vertex] (0) at (0, 0.15) {};
		\node [style=green vertex] (1) at (0, -0.15) {};
	\end{pgfonlayer}
	\begin{pgfonlayer}{edgelayer}
		\draw (0) to (1);
	\end{pgfonlayer}
\end{tikzpicture}}
\endpgfgraphicnamed = %
\beginpgfgraphicnamed{dashedbox}
\InputIfFileExists{dashedbox.tikz}{}{\input{./figures/dashedbox.tikz}}
\endpgfgraphicnamed
  \end{equation}
  In \cite{Coecke:2009aa} this structure is called a \emph{scaled}
  bialgebra.  The usual definition can be restored by imposing
  (\ref{eq:scalar-is-trivial}).  Space does not permit a full
  discussion of the scalars but note that equation
  (\ref{eq:scalar-is-trivial}) is not true in the standard model
  $\mathbb{CZ}_D$.  However, having belaboured the point that the
  scalars are needed, we henceforward omit them in the name of clarity
  -- they can always be restored if needed: see Backens
  \cite{Backens:2015aa}.
\end{remark}

\begin{definition}\label{def:hopf-alg}
  A bialgebra on $A$ is called a \emph{Hopf algebra} and if there
  exists $s:A \to A$, called the \emph{antipode}, satisfying the
  equation
  \begin{equation}\label{eq:hopf-law}\tag{H}
\beginpgfgraphicnamed{hopf-law}
\InputIfFileExists{hopf-law.tikz}{}{\input{./figures/hopf-law.tikz}}
\endpgfgraphicnamed \;= %
\beginpgfgraphicnamed{gr-zero}
\begin{tikzpicture}
	\begin{pgfonlayer}{nodelayer}
		\node [style=green vertex] (0) at (0, 0.5) {};
		\node [style=red vertex] (1) at (0, -0.5) {};
		\node [style=none] (2) at (0, -1) {};
		\node [style=none] (3) at (0, 1) {};
	\end{pgfonlayer}
	\begin{pgfonlayer}{edgelayer}
		\draw (2.center) to (1);
		\draw (0) to (3.center);
	\end{pgfonlayer}
\end{tikzpicture}}
\endpgfgraphicnamed
  \end{equation}
\end{definition}

\begin{definition}\label{def:antipode}
  Let $(\gdelta,\gepsilon,\rmu,\reta)$ be a
  Frobenius bialgebra as above; define the \emph{antipode} $s$ as 
  \[
  s = %
\beginpgfgraphicnamed{antipode}
\begin{tikzpicture}
	\begin{pgfonlayer}{nodelayer}
		\node [style=antipode] (0) at (0, 0) {};
		\node [style=none] (1) at (0, 0.5) {};
		\node [style=none] (2) at (0, -0.5) {};
	\end{pgfonlayer}
	\begin{pgfonlayer}{edgelayer}
		\draw (1.center) to (0);
		\draw (0) to (2.center);
	\end{pgfonlayer}
\end{tikzpicture}}
\endpgfgraphicnamed := %
\beginpgfgraphicnamed{gcap-rcup}
\InputIfFileExists{gcap-rcup.tikz}{}{\input{./figures/gcap-rcup.tikz}}
\endpgfgraphicnamed
  \]
\end{definition}

\begin{theorem}  \label{thm:hopf-iff-units-are-real}
  The morphisms $(\gdelta,\gepsilon,\rmu,\reta)$ form a Hopf algebra if and only
  if $\reta = (\repsilon)\gtrans$ and $\gepsilon = (\geta)\rtrans$, \ie
  \begin{equation}
    \label{eq:BAplus}\tag{+}
\beginpgfgraphicnamed{gunit-rreal}
\InputIfFileExists{gunit-rreal.tikz}{}{\input{./figures/gunit-rreal.tikz}}
\endpgfgraphicnamed = \gcounit
    \qquad\qquad
\beginpgfgraphicnamed{runit-greal}
\InputIfFileExists{runit-greal.tikz}{}{\input{./figures/runit-greal.tikz}}
\endpgfgraphicnamed = \runit
  \end{equation}
\end{theorem}

\begin{remark}\label{rem:classical-point-are-real}
  In the original paper on interacting quantum observables
  \cite{Coecke:2009aa} the condition ``\rdot-classical points are
  \rdot-real'' formed part of the definition of complementarity;
  equation (\ref{eq:BAplus}) is a weakening of this condition.
\end{remark}

Equation (\ref{eq:BAplus}) can be stated in purely Hopf algebraic
terms as
\[
\beginpgfgraphicnamed{antipode-gcounit}
\begin{tikzpicture}
	\begin{pgfonlayer}{nodelayer}
		\node [style=antipode] (0) at (0, 0) {};
		\node [style=none] (1) at (0, 0.5) {};
		\node [style=green vertex] (2) at (0, -0.5) {};
	\end{pgfonlayer}
	\begin{pgfonlayer}{edgelayer}
		\draw (1.center) to (0);
		\draw (0) to (2);
	\end{pgfonlayer}
\end{tikzpicture}}
\endpgfgraphicnamed = \gcounit
\qquad\qquad
\beginpgfgraphicnamed{antipode-redunit}
\begin{tikzpicture}
	\begin{pgfonlayer}{nodelayer}
		\node [style=antipode] (0) at (0, 0) {};
		\node [style=red vertex] (1) at (0, 0.5) {};
		\node [style=none] (2) at (0, -0.5) {};
	\end{pgfonlayer}
	\begin{pgfonlayer}{edgelayer}
		\draw (1) to (0);
		\draw (0) to (2.center);
	\end{pgfonlayer}
\end{tikzpicture}}
\endpgfgraphicnamed = \runit\;,
\]
however the given version emphasises that it is an interaction of
the red and green monoid structures, but not a complete distributive
law.  Indeed, as we shall see later, there is no general distributive
law of $\FA G$ over $\FA H$.  We are forced to define the PROP of
interacting Frobenius algebras as a quotient.

\begin{definition}\label{def:PROP-IF}
  Let $\IFA(G,H)$ be the PROP obtained quotienting $\FA G+\FA
  H$ by the equations (B+).  This gives a functor $\IFA : \Ab \times
  \Ab \to \dPROP$.
\end{definition}

Whenever the groups $G$ and $H$ are obvious or unimportant, we
abbreviate $\IFA(G,H)$ by \IFA.

\begin{example}\label{ex:stdmodel}
  The group algebras $\mathbb{CZ}_D$ described in
  Section~\ref{sec:standard-model} are
  $\IFA(\mathbb{Z}_D,\mathbb{Z}_D)$-algebras.  Indeed, the same group
  algebras are models of $\IFA(T^{D-1},T^{D-1})$, where 
  $T^n$ is the $n$-torus, \ie the $n$-fold product of circles.
  For $D=2$ this yields the usual model of the \zxcalculus.
\end{example}

\IFA contains two copies of the PROP of bialgebras: $\BA$ generated by
$(\gdelta,\rmu)$ and $\BA\op$ by $(\rdelta,\gmu)$.  By Theorem
\ref{thm:hopf-iff-units-are-real} $\IFA$ also contains two Hopf
algebra structures. Let \HA be the subcategory
generated by \BA and $s$, and define $\HA\op$ dually.  Note that we
have an isomorphism $\HA \isomorphism \HA\op$ via the dagger.


\begin{proposition}\label{prop:antipode-facts}
  Let $s$ be the antipode of a commutative Hopf algebra $H$; then
  \begin{enumerate}
  \item $s$ is the unique map satisfying \emph{(H)}; 
  \item $s$ is a bialgebra morphism; 
  \item $s\circ s = \id{}$ ;
  \item Let $K$ be a commutative Hopf algebra with antipode $s'$; then
    for any any bialgebra morphism $f : H \to K$ we have $f\circ s =
    s' \circ f$ 
  \end{enumerate}
\end{proposition}
These are standard properties (see \cite{street2007quantum}) which lead immediately to the following.

\begin{corollary}\label{cor:antipode-is-nice-more}  Let $s$ be as
  defined in \ref{def:antipode}; then:
  \begin{enumerate}
  \item $s$ is a self-adjoint unitary
  \item $s = s\gconj = s\rconj$
  \end{enumerate}
\end{corollary}

%

\begin{remark}\label{rem:same-s-opp-HA}
  Since the Hopf algebra is built out of \dg-Frobenius algebras, and
  $s = s^\dg$, we know that $s$ is also the antipode for the opposite
  bialgebra.
\end{remark}

Corollary \ref{cor:antipode-is-nice-more}  imply that the 
antipode commutes with all of the structure in of $\IFA(1,1)$.  This forces 
the two transpositions to interact in a variety of unexpected ways.

\begin{lemma}\label{lem:rg-equals-gr}
  For any $f: n\to m$ we have $f\grtrans = f\rgtrans$.
  \begin{proof}
    \[
    f\grtrans 
    = (s^{\otimes m}) \circ f \circ ({s^\dg}^{\otimes n})
    = ({s^\dg}^{\otimes m}) \circ f \circ ({s}^{\otimes n})
    =  f\rgtrans\;.
    \]
  \end{proof}
\end{lemma}

\begin{corollary}\label{cor:rg-props}
  \begin{enumerate}
  \item If $f$ commutes with $s$ then $f = f\rgtrans$;
  \item If $f$ commutes with $s$ then $f$ is \gdot-real iff it is
    \rdot-real;
  \item If $f$ is both \gdot-real and \rdot-real then it commutes with
    $s$.
  \end{enumerate}
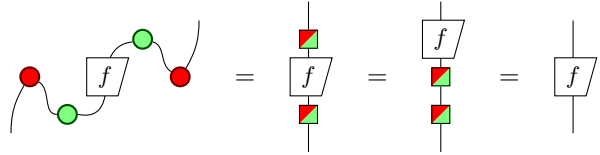
  \begin{proof} 1.
  \[
\beginpgfgraphicnamed{transpose-cor1}
\InputIfFileExists{transpose-cor1.tikz}{}{\input{./figures/transpose-cor1.tikz}}
\endpgfgraphicnamed \quad = \quad 
\beginpgfgraphicnamed{transpose-cor2}
\InputIfFileExists{transpose-cor2.tikz}{}{\input{./figures/transpose-cor2.tikz}}
\endpgfgraphicnamed \quad = \quad %
\beginpgfgraphicnamed{transpose-cor3}
\InputIfFileExists{transpose-cor3.tikz}{}{\input{./figures/transpose-cor3.tikz}}
\endpgfgraphicnamed \quad = \quad 
\beginpgfgraphicnamed{transpose-cor4}
\begin{tikzpicture}
	\begin{pgfonlayer}{nodelayer}
		\node [style=map] (0) at (0, -0) {$f$};
		\node [style=none] (1) at (0, -0.75) {};
		\node [style=none] (2) at (0, 0.75) {};
	\end{pgfonlayer}
	\begin{pgfonlayer}{edgelayer}
		\draw [style=simple] (2.center) to (0);
		\draw [style=simple] (0) to (1.center);
	\end{pgfonlayer}
\end{tikzpicture}}
\endpgfgraphicnamed
  \]

2. Suppose $f^{\dg} = f\gtrans$, then by the above $f = f\grtrans =
{f^{\dg}}{}\rtrans = f\rconj$
The converse holds by the same argument.

3. Suppose $f$ is \gdot-real and \rdot-real;  then $f=f\rgtrans =
sfs$ which gives the result by Corollary \ref{prop:antipode-facts}.3.
  \end{proof}
\end{corollary}

\begin{corollary}\label{cor:antipode-and-real-points}
  Suppose $k:0\to 1$ is \gdot-real, and let $h=\Lambda\rconj (k)$.  Then
  \begin{enumerate}
  \item $s\circ k = k\rconj$, and
  \item $s\circ h \circ s = h^\dg$.
  \end{enumerate}
\end{corollary}

\begin{proposition}\label{prop:IF-transpose-equals-dag}
  In \emph{$\IFA(1,1)$} $f\gtrans \!= f\rtrans \!= f^\dg$ for all morphisms $f$.
  \begin{proof}
    The generators of $\IFA(1,1)$ are real in their own colour, and as noted
    above $s$ commutes with all the generators of the PROP; hence the
    result follows by Corollary \ref{cor:rg-props}.
  \end{proof}
\end{proposition}


Given a pair of bialgebras $(A, \mu_A, \delta_A)$ and 
$(B,\mu_B, \delta_B)$, the collection of morphisms $A \to B$ 
becomes a monoid under the \emph{convolution product} where $f + g := \mu_B 
\circ (f \otimes g) \circ \delta_A$, and the unit is $\eta_A \circ \epsilon_B$.

In particular the endomorphisms of the bialgebra 
$(\gdelta,\gepsilon,\rmu,\reta)$ carry this monoid structure.

\begin{proposition}\label{prop:internal-ints}
  Let $f$ be a bialgebra morphism, and $s$ the antipode; then for all
  $g,h:A\to A$ we have:
  \begin{enumerate}
  \item $f \circ (g + h) = (f\circ g) + (f\circ h)\;$,
  \item $(g + h) \circ f = (g\circ f) + (h\circ f)\;$, and
  \item  $f + (f\circ s) = 0\;$. 
  \end{enumerate}
\end{proposition}
\begin{lemma}
  \label{lem:bialg-morphs-closed-under-conv}
  If $f$ and $g$ are bialgebra morphisms then so is $f+g$.
\end{lemma}

Hence the bialgebra morphisms of $(\gdelta,\gepsilon,\rmu,\reta)$ form
a unital ring $R$, with multiplication given by composition, and where the 
additive inverse is given by composing with $s$.  Accordingly we refer to the 
morphisms $\mathbf{n} \in
\mathbb{Z}$ as the \emph{internal integers}.

Define
$\mathbf{n}:A\to A$ by
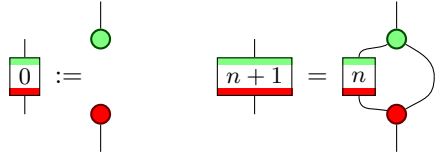
\[
\beginpgfgraphicnamed{grint-zero}
\begin{tikzpicture}
	\begin{pgfonlayer}{nodelayer}
		\node [style=grint] (0) at (0, 0) {$0$};
		\node [style=none] (1) at (0, 0.5) {};
		\node [style=none] (2) at (0, -0.5) {};
	\end{pgfonlayer}
	\begin{pgfonlayer}{edgelayer}
		\draw (1.center) to (0);
		\draw (0) to (2.center);
	\end{pgfonlayer}
\end{tikzpicture}}
\endpgfgraphicnamed \;:= %
\beginpgfgraphicnamed{gr-zero}
\begin{tikzpicture}
	\begin{pgfonlayer}{nodelayer}
		\node [style=green vertex] (0) at (0, 0.5) {};
		\node [style=red vertex] (1) at (0, -0.5) {};
		\node [style=none] (2) at (0, -1) {};
		\node [style=none] (3) at (0, 1) {};
	\end{pgfonlayer}
	\begin{pgfonlayer}{edgelayer}
		\draw (2.center) to (1);
		\draw (0) to (3.center);
	\end{pgfonlayer}
\end{tikzpicture}}
\endpgfgraphicnamed
\qquad\qquad
\beginpgfgraphicnamed{grint-nplus1}
\begin{tikzpicture}
	\begin{pgfonlayer}{nodelayer}
		\node [style=grint] (0) at (0, 0) {$n+1$};
		\node [style=none] (1) at (0, 0.5) {};
		\node [style=none] (2) at (0, -0.5) {};
	\end{pgfonlayer}
	\begin{pgfonlayer}{edgelayer}
		\draw (1.center) to (0);
		\draw (0) to (2.center);
	\end{pgfonlayer}
\end{tikzpicture}}
\endpgfgraphicnamed \;=\; %
\beginpgfgraphicnamed{grint-n-def}
\InputIfFileExists{grint-n-def.tikz}{}{\input{./figures/grint-n-def.tikz}}
\endpgfgraphicnamed
\]
for all $n \in \mathbb{N}$.  Applying the 
bialgebra law and the spider theorem respectively we have:
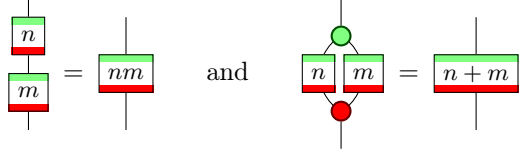
\[
\beginpgfgraphicnamed{grints1}
\begin{tikzpicture}
	\begin{pgfonlayer}{nodelayer}
		\node [style=none] (0) at (0, 1) {};
		\node [style=none] (1) at (0, -0.75) {};
		\node [style=grint] (2) at (0, 0.5) {$n$};
		\node [style=grint] (3) at (0, -0.25) {$m$};
	\end{pgfonlayer}
	\begin{pgfonlayer}{edgelayer}
		\draw (0.center) to (2);
		\draw (2) to (3);
		\draw (3) to (1.center);
	\end{pgfonlayer}
\end{tikzpicture}}
\endpgfgraphicnamed \;=\; %
\beginpgfgraphicnamed{grints2}
\begin{tikzpicture}
	\begin{pgfonlayer}{nodelayer}
		\node [style=none] (0) at (0, 0.75) {};
		\node [style=none] (1) at (0, -0.75) {};
		\node [style=grint] (2) at (0, 0) {$nm$};
	\end{pgfonlayer}
	\begin{pgfonlayer}{edgelayer}
		\draw (0.center) to (2);
		\draw (1.center) to (2);
	\end{pgfonlayer}
\end{tikzpicture}}
\endpgfgraphicnamed
\qquad\text{and}\qquad
\beginpgfgraphicnamed{grints4}
\InputIfFileExists{grints4.tikz}{}{\input{./figures/grints4.tikz}}
\endpgfgraphicnamed  \;=\; %
\beginpgfgraphicnamed{grints3}
\begin{tikzpicture}
	\begin{pgfonlayer}{nodelayer}
		\node [style=none] (0) at (0, 0.75) {};
		\node [style=none] (1) at (0, -0.75) {};
		\node [style=grint] (2) at (0, 0) {$n+m$};
	\end{pgfonlayer}
	\begin{pgfonlayer}{edgelayer}
		\draw (0.center) to (2);
		\draw (2) to (1.center);
	\end{pgfonlayer}
\end{tikzpicture}}
\endpgfgraphicnamed
\]
Further, $\mathbf{n}$ is a bialgebra morphism for
$(\gdelta,\gepsilon,\rmu,\reta)$.

\begin{example}\label{ex:internal-ints-Zthree}
  In the group algebra $\mathbb{CZ}_3$ the internal integers are
  given by the following matrices:
\[
\mathbf{0} = \begin{pmatrix}
  1&0&0\\1&0&0\\1&0&0
\end{pmatrix} \quad
\mathbf{1} = \begin{pmatrix}
  1&0&0\\0&1&0\\0&0&1
\end{pmatrix} \quad
\mathbf{2} = \begin{pmatrix}
  1&0&0\\0&0&1\\0&1&0
\end{pmatrix}
\]
There are no others;  see Section \ref{sec:finite-dimension} for discussion.
\end{example}

\begin{lemma}\label{lem:phases-are-not-bialg}
  Let $\alpha$ be a phase of either colour in \emph{$\IFA(G,H)$}; then
  $\alpha$ is bialgebra morphism iff $\alpha = \id{}$.
\end{lemma}

Since the non-trivial phases can never be bialgebra morphisms, we restrict our
attention to $\IFA(1,1)$ for the rest of this section.

In any PROP, given a monoid $(\mu,\eta)$ on 1, a monoid on 2 can be
defined using the tensor:
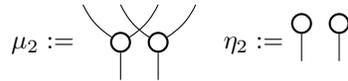
\[
\mu_2 := %
\beginpgfgraphicnamed{tensor-monoid1}
\InputIfFileExists{tensor-monoid1.tikz}{}{\input{./figures/tensor-monoid1.tikz}}
\endpgfgraphicnamed \quad \eta_2 := %
\beginpgfgraphicnamed{tensor-monoid2}
\InputIfFileExists{tensor-monoid2.tikz}{}{\input{./figures/tensor-monoid2.tikz}}
\endpgfgraphicnamed
\]
and similarly for a comonoid $(\delta,\epsilon)$.  It is easy to check
that if $(\delta,\epsilon,\mu,\eta)$ is a bi-, Hopf, or Frobenius
algebra then so is $(\delta_2,\epsilon_2,\mu_2,\eta_2)$.  Continuing
in the same way, there is a Hopf algebra on every object $n$ of \HA.
Therefore all the preceding discussion applies equally well to
bialgebra morphisms $n\to m$ for any $n$ and $m$.  In particular all
the generators of \HA are bialgebra morphisms, which yields:

\begin{lemma}
Every morphism in \emph{$\HA$} is a bialgebra homomorphism for $(\gdelta, 
\rmu)$.
\end{lemma}

Thanks to \dg-duality, any bialgebra morphism for $f\in\HA$ gives a
bialgebra morphism $f^\dg\in\HA\op$, so we also
have an isomorphic opposite ring $R\op$, complete with opposite
integers $\mathbf{n}^\dg$.  None of the equations of \IFA forces these
structures to interact: there is no commutation of $\gdelta$ and
$\rdelta$ for example.  However, if we restrict attention to the
invertible morphisms of $\IFA(1,1)$ we can make some progress.


\begin{lemma}\label{lem:inv-implies-bialg}
If $f$ is an invertible bialgebra morphism then $f^{-1}$ is also a bialgebra 
morphism.
\end{lemma}

\begin{lemma}\label{lem:invertible-integers}
  Let $f: n \to n$ be a bialgebra morphism, and suppose that $f'\circ f = 
f\circ f' =
  \id{}$ for some morphism $f:n\to n$, which is both \gdot-real and
  \rdot-real; then $f' = f^\dg$.  

\begin{proof} By uniqueness of inverses in monoids it is enough to 
show that $(-f') + f^\dg = 0$.
    
   \[
\beginpgfgraphicnamed{lem-inv-int1}
\InputIfFileExists{lem-inv-int1.tikz}{}{\input{./figures/lem-inv-int1.tikz}}
\endpgfgraphicnamed \quad = \, \,  %
\beginpgfgraphicnamed{lem-inv-int2}
\InputIfFileExists{lem-inv-int2.tikz}{}{\input{./figures/lem-inv-int2.tikz}}
\endpgfgraphicnamed\quad  = \quad  
\beginpgfgraphicnamed{lem-inv-int3}
\InputIfFileExists{lem-inv-int3.tikz}{}{\input{./figures/lem-inv-int3.tikz}}
\endpgfgraphicnamed 
   \]
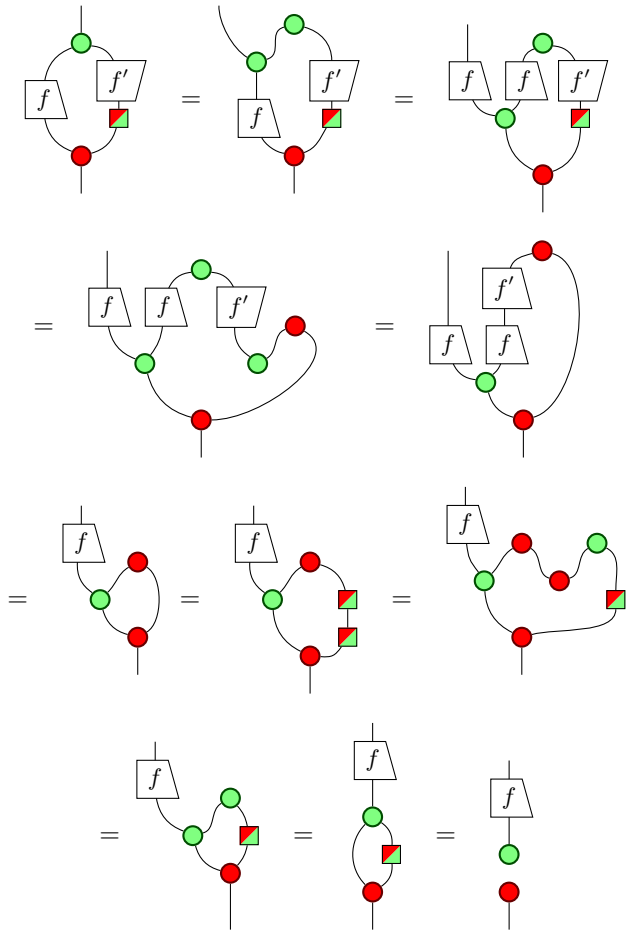
\[
  =\quad %
\beginpgfgraphicnamed{lem-inv-int3b}
\InputIfFileExists{lem-inv-int3b.tikz}{}{\input{./figures/lem-inv-int3b.tikz}}
\endpgfgraphicnamed\ \  =\quad   %
\beginpgfgraphicnamed{lem-inv-int4}
\InputIfFileExists{lem-inv-int4.tikz}{}{\input{./figures/lem-inv-int4.tikz}}
\endpgfgraphicnamed  
\]
\[
= \quad  %
\beginpgfgraphicnamed{lem-inv-int5}
\InputIfFileExists{lem-inv-int5.tikz}{}{\input{./figures/lem-inv-int5.tikz}}
\endpgfgraphicnamed\, \,  =\quad  %
\beginpgfgraphicnamed{lem-inv-int6}
\InputIfFileExists{lem-inv-int6.tikz}{}{\input{./figures/lem-inv-int6.tikz}}
\endpgfgraphicnamed\quad  
=\quad   %
\beginpgfgraphicnamed{lem-inv-int7}
\InputIfFileExists{lem-inv-int7.tikz}{}{\input{./figures/lem-inv-int7.tikz}}
\endpgfgraphicnamed 
\]
\[
 =\, \, %
\beginpgfgraphicnamed{lem-inv-int8}
\InputIfFileExists{lem-inv-int8.tikz}{}{\input{./figures/lem-inv-int8.tikz}}
\endpgfgraphicnamed \quad = \quad  %
\beginpgfgraphicnamed{lem-inv-int9}
\InputIfFileExists{lem-inv-int9.tikz}{}{\input{./figures/lem-inv-int9.tikz}}
\endpgfgraphicnamed \quad  
=\quad   %
\beginpgfgraphicnamed{lem-inv-int10}
\InputIfFileExists{lem-inv-int10.tikz}{}{\input{./figures/lem-inv-int10.tikz}}
\endpgfgraphicnamed 
\]
by a similar argument we obtain
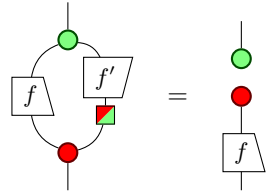
\[
\beginpgfgraphicnamed{lem-inv-int1}
\InputIfFileExists{lem-inv-int1.tikz}{}{\input{./figures/lem-inv-int1.tikz}}
\endpgfgraphicnamed \quad = \quad  %
\beginpgfgraphicnamed{lem-inv-int11}
\InputIfFileExists{lem-inv-int11.tikz}{}{\input{./figures/lem-inv-int11.tikz}}
\endpgfgraphicnamed
\]
It is easy to see the following
\[
\beginpgfgraphicnamed{lem-inv-int10}
\InputIfFileExists{lem-inv-int10.tikz}{}{\input{./figures/lem-inv-int10.tikz}}
\endpgfgraphicnamed \quad = \quad  %
\beginpgfgraphicnamed{lem-inv-int11}
\InputIfFileExists{lem-inv-int11.tikz}{}{\input{./figures/lem-inv-int11.tikz}}
\endpgfgraphicnamed \quad 
\Rightarrow \quad %
\beginpgfgraphicnamed{lem-inv-int12}
\begin{tikzpicture}
	\begin{pgfonlayer}{nodelayer}
		\node [style=green vertex] (0) at (0, -0.5) {};
		\node [style=mapdag] (1) at (0, 0.25) {$f$};
		\node [style=none] (2) at (0, 0.75) {};
	\end{pgfonlayer}
	\begin{pgfonlayer}{edgelayer}
		\draw [style=simple] (1) to (2.center);
		\draw [style=simple] (0) to (1);
	\end{pgfonlayer}
\end{tikzpicture}}
\endpgfgraphicnamed \quad = \quad 
\beginpgfgraphicnamed{lem-inv-int12b}
\begin{tikzpicture}
	\begin{pgfonlayer}{nodelayer}
		\node [style=green vertex] (0) at (0, -0.25) {};
		\node [style=none] (1) at (0, 0.5) {};
	\end{pgfonlayer}
	\begin{pgfonlayer}{edgelayer}
		\draw [style=simple] (1.center) to (0);
	\end{pgfonlayer}
\end{tikzpicture}}
\endpgfgraphicnamed \quad
\]
and hence, $(-f') + f^\dg = 0$ as required.
  \end{proof}
\end{lemma}

\begin{theorem}\label{thm:invertible-HA-implies-unitary}
Let \emph{$f \in \IFA$} be an invertible morphism;  if \emph{$f\in\HA$} then $f$
is unitary.
\begin{proof}
  By proposition \ref{prop:IF-transpose-equals-dag}, all morphisms in
  $\IFA(1,1)$ are $\gdot$-real and $\rdot$-real, so in particular
  $f^{-1}$ is.  Since $f \in \HA$ it is a bialgebra morphism.  Hence
  the result follows from Lemma \ref{lem:invertible-integers}.
\end{proof}
\end{theorem}

Combining the preceding result with Lemma \ref{lem:inv-implies-bialg}
shows that the invertible elements of the rings $R$ and $R\op$ must
coincide.  However, we can do better by appealing to this theorem of
Bonchi et al:

\begin{theorem}{\cite[Prop. 3.7]{Bonchi2014a}}
\label{thm:HA-equiv-matrices}
  Let $\mathrm{Mat}R$ denote the category of matrices valued in the
  ring $R$; then \emph{$\HA \simeq \mathrm{Mat} R$} is an isomorphism  of
  PROPS.
\end{theorem}

\begin{theorem}\label{thm:HA-units-coincide}
  Let \emph{$f \in \IFA$} be invertible; then \emph{$f\in \HA$} iff 
\emph{$f\in\HA\op$}
  \begin{proof}
    By Theorem \ref{thm:invertible-HA-implies-unitary} \emph{$f^{-1} = f^\dg
    \in \HA\op$}.  However by Theorem \ref{thm:HA-equiv-matrices} the
    inverse of an $R$-valued matrix is again an $R$-valued matrix,
    hence $f^{-1} \in \HA$
  \end{proof}
\end{theorem}

\begin{remark}\label{rem:actually-no-invertible-things-in-R}
  \emph{A priori} there are no invertible elements of $R$ other than
  the identity and the antipode in $\IFA(1,1)$.  However, when we
  consider the finite dimensional collapse of $\IFA$ in Section
  \ref{sec:finite-dimension} the this will no longer be true.  (By
  Lemma \ref{lem:phases-are-not-bialg} it suffices to consider
  $\IFA(1,1)$.)
\end{remark}

\section{Set-like elements and classical maps}
\label{sec:set-like-elements}

Recall that $\psi:I \to A$ is called \emph{set-like} (also called
\emph{group-like} or \emph{classical}) for the
coalgebra $\delta:A \to A \otimes A$ if $\delta(\psi) = \psi \otimes
\psi$.  Set-like elements of a \dg-SCFA correspond to the eigenstates of
an observable in quantum mechanics, hence they are of great importance
for applications.  They have many useful properties, which we now
explore.


The following is standard; see \cite{Sweedler1969Hopf-Algebras}.
\begin{lemma}\label{lem:setlike-props}
Let $(\mu, \delta, \eta, \epsilon)$ be a bialgebra; then:
\begin{enumerate}
\item $\eta$ is set-like.
\item If  $\psi$ and $\varphi$ are set-like then $\mu(\psi
\otimes \varphi)$ is set-like.
\item If this bialgebra is a Hopf algebra with antipode $s$ then
  $\mu(\psi \otimes (s\circ\psi)) = \eta$.
\end{enumerate}
Hence the set-like elements form a monoid for every bialgebra and a
group for every Hopf algebra.
\end{lemma}


In an \IFA-algebra we have two coalgebras, and hence
two ways to be set-like.  Call an element $\rdot$-\emph{set-like} if it
is set-like for $\rdelta$, and similarly for $\gdelta$.

\begin{lemma}\label{lem:setlike-implies-unbiased}
  Let $k$ be a \gdot-set-like element in some \emph{\IFA}-algebra; then $k$
  is \rdot-unbiased iff it is \gdot-real.
  \begin{proof}
    If $k$ is \gdot-real, we have $s\circ k = k\rconj$, hence by Lemma
    \ref{lem:setlike-props}.3 we have $\reta = \rmu(k \otimes
    (s\circ k)) = \rmu(k \otimes k\rconj)$ which implies $k$ is
    \rdot-unbiased by Lemma \ref{lem:phases-and-unb-points}.  Conversely, if
    $k$ is \rdot-unbiased then, by Lemma \ref{lem:setlike-props} and
    the uniqueness of inverses, we have $s\circ k = k\rconj$ from which
    $k^\dg = k\gtrans$ follows.
  \end{proof}
\end{lemma}

\begin{remark}\label{rem:classical-points-are-real-again}
  Again, this clarifies the ``classical points are real'' assumption
  of \cite{Coecke:2009aa} -- in that work \gdot-set-like elements are
  separately assumed to be both \gdot-real and \rdot-unbiased.
  In vector space models, $k$ being \gdot-real means that all
    its matrix entries are real when written in the orthonormal basis
    defining \gdot; this seems a very natural property to demand of
    the basis vectors themselves!  However there is no a priori reason
    why this should coincide with being \rdot-unbiased; it is
    surprising that these properties are axiomatically equivalent.
\end{remark}

\begin{corollary}\label{cor:g-setlike-implies-r-unbiased}
  Let $\rdelta$ and $\gdelta$ be a pair of interacting \dg-SCFAs and
  suppose that the \whitedot-set-like elements are also
  \whitedot-real, for $\whitedot \in \{\gdot, \rdot\}$; then
  \rdot-set-like elements form a subgroup of the \gdot-unbiased points
  and vice versa.
\end{corollary}

By Lemma \ref{lem:phases-and-unb-points} we know that each
$\rdot$-unbiased point $\alpha$ determines an $\rdot$-phase
$\Lambda\rconj(\alpha)$.  Phases that are constructed from
$\gdot$-set-like points are called $\gdot$-\emph{classical}.  It
follows immediately from Corollary \ref{cor:g-setlike-implies-r-unbiased}
that the $\gdot$-classical maps form a subgroup of the $\rdot$-phases.

\begin{lemma}\label{lem:no-classical-phases}
No \gdot-phase can also be \gdot-classical.
\begin{proof}
  If  $k:I \to A$ is $\gdot$-set-like then $\Lambda\gconj (k) = k\circ
  k\gtrans$, which is a projector, hence not unitary and therefore not a phase.
\end{proof}
\end{lemma}

Bearing Lemmas \ref{lem:setlike-implies-unbiased} and
  \ref{lem:no-classical-phases} in mind we wish to consider
  interacting Frobenius algebras where we have a given subgroup of
  phases for one colour which are classical for the other.

\begin{definition}\label{def:IFK}
  Let $G_K$ and $H_K$ be subgroups of abelian groups $G$ and $H$
  respectively.  Define the \dg-PROP $\IFK(G\geq G_K ,H\geq H_K)$ as
  the quotient obtained by imposing on $\IFA(G,H)$ the equations
\begin{equation}\label{IFKequations}\tag{IFK}
\beginpgfgraphicnamed{IFKeqns1}
\InputIfFileExists{IFKeqns1.tikz}{}{\input{./figures/IFKeqns1.tikz}}
\endpgfgraphicnamed \ = \ %
\beginpgfgraphicnamed{IFKeqns3}
\InputIfFileExists{IFKeqns3.tikz}{}{\input{./figures/IFKeqns3.tikz}}
\endpgfgraphicnamed \quad \quad \quad \quad 
\beginpgfgraphicnamed{IFKeqns2}
\InputIfFileExists{IFKeqns2.tikz}{}{\input{./figures/IFKeqns2.tikz}}
\endpgfgraphicnamed \ =  \ %
\beginpgfgraphicnamed{IFKeqns4}
\InputIfFileExists{IFKeqns4.tikz}{}{\input{./figures/IFKeqns4.tikz}}
\endpgfgraphicnamed 
\end{equation}
for each $g \in G_K$ and each $h\in H_K$
\end{definition}

\begin{remark}\label{rem:Hk-Gk-not-always-iso}
  Note that in complex Hilbert space models we must have $H_K \isomorphism 
  G_K$ but there are concrete models in which this is not the
  case. For example, consider the groups algebra of $\mathbb{Z}_4$
  over the reals. The $\gdot$-set-like elements correspond with the
  elements of $\mathbb{Z}_4$ but by proposition
  \ref{lem:characters-are-set-like-for-group-algs} the
  $\rdot$-set-like elements correspond with the group homomorphisms
  $\mathbb{Z}_4 \to \mathbb{R}^{\times}$, of which there are only
  two. Hence as groups the set-like elements for the respective
  colours are not isomorphic.
\end{remark}

\begin{lemma}\label{lem:classical-commutes-with-structure}
  If $h:1\to 1$ is \gdot-classical then it commutes with
  $\gdelta$ and $\gmu$.
\end{lemma}

%

In the $\dg$-PROP $\IFA(G,H)$, by definition the only phases for the
respective colours are $G$ and $H$. However, the presence of classical
maps in $\IFK(G\geq G_K, H \geq H_K)$ changes this.

\begin{theorem}
Let $\alpha$ be a $\gdot$-phase and $k$ a $\gdot$-classical map then
$k\circ \alpha \circ k^\dg$
is a $\gdot$-phase
\begin{proof}  By its construction $k\circ \alpha \circ
  k^\dg$ is evidently unitary.  We need to show that it is a pre-phase.
Consider
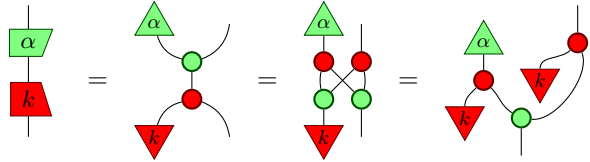
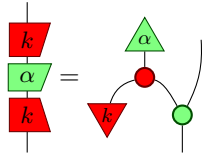
\[
\beginpgfgraphicnamed{haction1}
\InputIfFileExists{haction1.tikz}{}{\input{./figures/haction1.tikz}}
\endpgfgraphicnamed \quad = \ \ %
\beginpgfgraphicnamed{haction2}
\InputIfFileExists{haction2.tikz}{}{\input{./figures/haction2.tikz}}
\endpgfgraphicnamed \ \ = \ \ %
\beginpgfgraphicnamed{haction3}
\InputIfFileExists{haction3.tikz}{}{\input{./figures/haction3.tikz}}
\endpgfgraphicnamed 
\ \ = \ \ %
\beginpgfgraphicnamed{haction6}
\InputIfFileExists{haction6.tikz}{}{\input{./figures/haction6.tikz}}
\endpgfgraphicnamed
\]
hence we have
\[
\beginpgfgraphicnamed{hactionstatement}
\InputIfFileExists{hactionstatement.tikz}{}{\input{./figures/hactionstatement.tikz}}
\endpgfgraphicnamed = %
\beginpgfgraphicnamed{haction4}
\InputIfFileExists{haction4.tikz}{}{\input{./figures/haction4.tikz}}
\endpgfgraphicnamed
\]
as required.
\end{proof}
\end{theorem}
While Lemma \ref{lem:no-classical-phases} tells us that the \gdot-phases and 
\gdot-classical maps are disjoint as groups, there is a degree of interaction 
between \gdot-phases and \gdot-classical maps: the classical maps act
on the phase group, to produce new phases.

\begin{lemma}\label{lem:classical-act-on-phases}
For $H_K$ the group of $\gdot$-set-like elements and $\Phi\gconj$ the group
of $\gdot$-phases there is a group action $\bullet : H_K \times \Phi\gconj 
\to 
\Phi\gconj$.
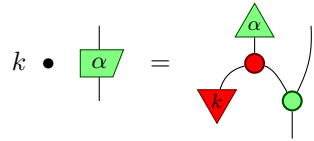
\[
k \ \bullet \ \ %
\beginpgfgraphicnamed{alpha-phase}
\begin{tikzpicture}
	\begin{pgfonlayer}{nodelayer}
		\node [style=green map] (0) at (0, -0) {$\alpha$};
		\node [style=none] (1) at (0, -0.5) {};
		\node [style=none] (2) at (0, 0.5) {};
	\end{pgfonlayer}
	\begin{pgfonlayer}{edgelayer}
		\draw [style=simple] (0) to (2.center);
		\draw [style=simple] (0) to (1.center);
	\end{pgfonlayer}
\end{tikzpicture}}
\endpgfgraphicnamed \ \ = \ \ %
\beginpgfgraphicnamed{haction4}
\InputIfFileExists{haction4.tikz}{}{\input{./figures/haction4.tikz}}
\endpgfgraphicnamed
\]
for $\alpha \in \Phi\gconj$ and $k \in H_K$
\end{lemma}

\begin{theorem}\label{thm:semi-direct-product}
The set of morphisms obtained by freely composing \gdot-phases and 
\gdot-classical maps is a group isomorphic to $\Phi\gconj 
\rtimes_{\varphi}~H_K$ the (outer) semidirect product of $\Phi\gconj$ and $H_K$ 
over the action~ $\bullet$.
\end{theorem}


We end this section with an important lemma relating the ring
structure and the classical maps.
  \begin{lemma}\label{lem:integers-and-classical-pts}
    Let $\mathbf{n} \in R$ be an internal integer, and $k:A \to A$ be
    a $\gdot$-classical map then $\mathbf{n} \circ k=k^n \circ
    \mathbf{n}$.
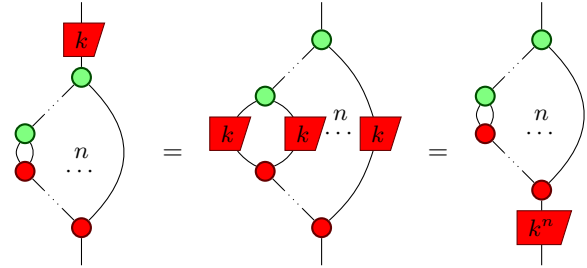
\begin{proof}
\[ 
\beginpgfgraphicnamed{akproof1}
\InputIfFileExists{akproof1.tikz}{}{\input{./figures/akproof1.tikz}}
\endpgfgraphicnamed \ \ = \ \ 
\beginpgfgraphicnamed{akproof2}
\InputIfFileExists{akproof2.tikz}{}{\input{./figures/akproof2.tikz}}
\endpgfgraphicnamed \ \ = \ \ %
\beginpgfgraphicnamed{akproof3}
\InputIfFileExists{akproof3.tikz}{}{\input{./figures/akproof3.tikz}}
\endpgfgraphicnamed 
\]
\end{proof}
\end{lemma}

\section{Collapse to finite dimension}
\label{sec:finite-dimension}

It is well known that if a vector space $A$ supports a Frobenius
algebra then $A$ must be finite dimensional.  As the last part of our
story we incorporate this fact into our axiomatic framework.  Recall that in a monoidal category an object $A$ is
said to have \emph{enough points} if, for all morphisms $f,g:A\to B$, we
have
\[
(\forall x : I\to A, \ \ fx = gx) \Rightarrow f = g\;.
\]
In vector spaces an even stronger form of extensionality is present:
two linear maps are equal if they agree on all elements of a basis for
the space.  Further, we have the following:

\begin{lemma}\label{lem:set-like-lin-indep}
  Let $\mathbbm{k}$ be a field, $A$ a $\mathbbm{k}$-vector space, and
  $\delta: A \to A \otimes A$ a coalgebra. The set-like elements of
  $\delta$ are linearly independent.
\begin{proof}
See \cite[Proposition 7.2]{street2007quantum}.
\end{proof}
\end{lemma}

This motivates the following definition.

\begin{definition}\label{def:enough-set-like-points}
  Let $A$ be an $\IFK(G\geq G_K, H \geq H_K)$-algebra; then $A$ has
  \emph{enough \gdot-set-like elements} if
\[(\forall k \in H_K \ : \ f \circ k \circ \reta = g \circ k \circ \reta)
  \Rightarrow f = g
\] 
holds for all $f,g:A\to B$.
\end{definition}

By lemma \ref{lem:set-like-lin-indep} it follows that an \IFK-algebra
in $\textbf{Vect}_{\mathbbm{k}}$ has enough \gdot-set-like elements iff the
\gdot-set-like elements form a basis.  This suffices to determine 
\gdelta uniquely, while the group structure of $H_K$ determines \rmu,
and the whole thing is just the group algebra $\mathbbm{k}H_K$; \cf Section
\ref{sec:standard-model}.  The dimension of the underlying vector
space $A$ is then $\sizeof{H_K}$.



Since arguments that depend on having enough set-like points are
quite common in the categorical quantum mechanics literature, we impose
this condition to define the final PROP of this paper.  Note, per
Remark \ref{rem:Hk-Gk-not-always-iso}, asking for both $G_K$ and $H_K$
to be enough points is too strong, so we just pick one.

\begin{definition}\label{def:IFKd}
  Let $\IFKd(G\geq G_K, H \geq H_K)$ denote the \dg-PROP obtained from
  $\IFK(G\geq G_K, H \geq H_K)$ by imposing the condition of enough
  \gdot-set-like elements from Def.~\ref{def:enough-set-like-points}.
\end{definition}

Recall that the exponent of a finite group is the least non-zero $n$ such
that $g^n = 1$ for all $g$.

\begin{theorem}\label{thm:ring-collapse-to-exponent}
  Suppose $H_K$ is finite, and let $d$ be its exponent; then in \IFKd
  the internal integers are the finite ring $\mathbb{Z}_d$
\begin{proof}
Applying lemma \ref{lem:integers-and-classical-pts} we have 
\[
\beginpgfgraphicnamed{kdproof1}
\begin{tikzpicture}
	\begin{pgfonlayer}{nodelayer}
		\node [style=red map] (0) at (0, 0.25) {$k$};
		\node [style=grint] (1) at (0, -0.5) {$d+1$};
		\node [style=red vertex] (2) at (0, 0.75) {};
		\node [style=none] (3) at (0, -1) {};
	\end{pgfonlayer}
	\begin{pgfonlayer}{edgelayer}
		\draw (0) to (1);
		\draw (2) to (0);
		\draw (3.center) to (1);
	\end{pgfonlayer}
\end{tikzpicture}}
\endpgfgraphicnamed = \ 
\beginpgfgraphicnamed{kdproof2}
\begin{tikzpicture}
	\begin{pgfonlayer}{nodelayer}
		\node [style=grint] (0) at (0, 0.25) {$d+1$};
		\node [style=red map] (1) at (0, -0.5) {$k^{d+1}$};
		\node [style=red vertex] (2) at (0, 0.75) {};
		\node [style=none] (3) at (0, -1) {};
	\end{pgfonlayer}
	\begin{pgfonlayer}{edgelayer}
		\draw (0) to (1);
		\draw (2) to (0);
		\draw (3.center) to (1);
	\end{pgfonlayer}
\end{tikzpicture}}
\endpgfgraphicnamed = \ 
\beginpgfgraphicnamed{kdproof3}
\begin{tikzpicture}
	\begin{pgfonlayer}{nodelayer}
		\node [style=red vertex] (0) at (0, 0.5) {};
		\node [style=red map] (1) at (0, -0.25) {$k^{d+1}$};
		\node [style=none] (2) at (0, -0.75) {};
	\end{pgfonlayer}
	\begin{pgfonlayer}{edgelayer}
		\draw (0) to (1);
		\draw (2.center) to (1);
	\end{pgfonlayer}
\end{tikzpicture}}
\endpgfgraphicnamed \ = \ 
\beginpgfgraphicnamed{kdproof4}
\begin{tikzpicture}
	\begin{pgfonlayer}{nodelayer}
		\node [style=red vertex] (0) at (0, 0.5) {};
		\node [style=red map] (1) at (0, 0) {$k$};
		\node [style=none] (2) at (0, -0.5) {};
	\end{pgfonlayer}
	\begin{pgfonlayer}{edgelayer}
		\draw (0) to (1);
		\draw (2.center) to (1);
	\end{pgfonlayer}
\end{tikzpicture}}
\endpgfgraphicnamed, 
\]
for each $k \in H_k$.  Since $H_K$ is enough points, $\mathbf{d+1} =
\id{}$, from whence $\mathbf{n} = \mathbf{n+d}$ for all internal integers
$\mathbf{n} \in R$.
\end{proof}
\end{theorem}


The following observation follows directly from Theorem 
\ref{thm:ring-collapse-to-exponent} and Lemma
\ref{lem:invertible-integers}.
\begin{corollary}\label{cor:primality-gives-unitarity}
If $n \in \mathbb{Z}_d$ has a mutliplicative inverse then $\textbf{n} \in R$ is 
unitary. In particular, if the group $H_K$ has prime exponent then every 
non-zero $\textbf{n} \in R$ is unitary.
\end{corollary}

\begin{theorem}\label{thm:primality-gives-distribution}
Let $d$ be the exponent of $H_K$ and $n\in R$ then the following are 
equivalent:
\begin{enumerate}
\item $d$ and $n$ are coprime.
\item 
$%
\beginpgfgraphicnamed{coprime1a}
\InputIfFileExists{coprime1a.tikz}{}{\input{./figures/coprime1a.tikz}}
\endpgfgraphicnamed = 
\beginpgfgraphicnamed{coprime1b}
\InputIfFileExists{coprime1b.tikz}{}{\input{./figures/coprime1b.tikz}}
\endpgfgraphicnamed$  \quad\text{and}  \quad
$\ \ %
\beginpgfgraphicnamed{coprime1c}
\begin{tikzpicture}
	\begin{pgfonlayer}{nodelayer}
		\node [style=none] (0) at (0, 0.5) {};
		\node [style=red vertex] (1) at (0, -0.5) {};
		\node [style=grint] (2) at (0, 0) {$n$};
	\end{pgfonlayer}
	\begin{pgfonlayer}{edgelayer}
		\draw (0.center) to (2);
		\draw (2) to (1);
	\end{pgfonlayer}
\end{tikzpicture}}
\endpgfgraphicnamed = 
\beginpgfgraphicnamed{coprime1d}
\begin{tikzpicture}
	\begin{pgfonlayer}{nodelayer}
		\node [style=none] (0) at (0, 0.25) {};
		\node [style=red vertex] (1) at (0, -0.25) {};
	\end{pgfonlayer}
	\begin{pgfonlayer}{edgelayer}
		\draw (0.center) to (1);
	\end{pgfonlayer}
\end{tikzpicture}}
\endpgfgraphicnamed$
\item $%
\beginpgfgraphicnamed{coprime2a}
\InputIfFileExists{coprime2a.tikz}{}{\input{./figures/coprime2a.tikz}}
\endpgfgraphicnamed = 
\beginpgfgraphicnamed{coprime2b}
\InputIfFileExists{coprime2b.tikz}{}{\input{./figures/coprime2b.tikz}}
\endpgfgraphicnamed$ \quad \text{and} \quad
$\ \ %
\beginpgfgraphicnamed{coprime2c}
\begin{tikzpicture}
	\begin{pgfonlayer}{nodelayer}
		\node [style=none] (0) at (0, -0.5) {};
		\node [style=green vertex] (1) at (0, 0.5) {};
		\node [style=grint] (2) at (0, 0) {$n$};
	\end{pgfonlayer}
	\begin{pgfonlayer}{edgelayer}
		\draw (0.center) to (2);
		\draw (2) to (1);
	\end{pgfonlayer}
\end{tikzpicture}}
\endpgfgraphicnamed = 
\beginpgfgraphicnamed{coprime2d}
\begin{tikzpicture}
	\begin{pgfonlayer}{nodelayer}
		\node [style=none] (0) at (0, -0.25) {};
		\node [style=green vertex] (1) at (0, 0.25) {};
	\end{pgfonlayer}
	\begin{pgfonlayer}{edgelayer}
		\draw (0.center) to (1);
	\end{pgfonlayer}
\end{tikzpicture}}
\endpgfgraphicnamed$
\end{enumerate}
\begin{proof}
  (1) $\Rightarrow$ (2) and (3): The integers $n$ and $d$ are coprime
  iff $n$ is invertible in $\mathbb{Z}_d$, in which case $\mathbf{n}
  \in R$ is invertible.  By Theorem
  \ref{thm:HA-units-coincide}, the invertible members of
  $R$ and $R\op$ coincide.  Since $\mathbf{n} \in R\op$ it is a
  bialgebra morphism for $\BA\op$ which implies both (2) and (3).  

(2)  $\Rightarrow$ (1):  Since $\mathbf{n}\in R$ it commutes with
$(\rmu, \reta)$, by assumption it commutes with $(\rdelta,\repsilon)$;
hence it is a Frobenius algebra morphism, and by
Lemma~\ref{lem:frob-morphisms-are-iso} it is invertible in $R$; hence
$n$ is coprime to $d$.

(3) $\Rightarrow$ (1) follows by the same argument as above.
\end{proof}
\end{theorem}

In the case where $H_K$ has prime exponent, this theorem demonstrates
the kind distributive equation between \HA and $\HA\op$ 
conspicuously absent from Section~\ref{sec:inter-frob-algebr}.  Indeed
equations of this type take part in the distributive law 
used to construct the PROP $\mathbb{IH}^w_R$ in \cite{Bonchi2014a}.
However, as we now see, such a law is not possible in our more general
setting.

\begin{theorem}\label{thm:no-dist-law}
  There is no distributive law of PROPs 
  \[
  \tau: \emph{\FA} G ; \emph{\FA} H \to \emph{\FA} H ; \emph{\FA} G
  \]
  which gives rise to \emph{$\IFA(G,H)$}.
  \begin{proof}
    If such distributive law exists then for every composable pair
    \[
    n \rTo^{g_1} l \rTo^{h_1} m
    \]
    in $\FA G ; \FA H$, there must exist an equal composable pair
    \[
    n \rTo^{h'} k \rTo^{g'} m
    \]
    in $\FA H ; \FA G$.  Consider the case $n = l = m = 1$; then $g_1$
    and $h_1$ are just group elements from $G$ and $H$ respectively.  Applying the Generalised Spider theorem in
    $\FA H$ and $\FA G $ separately, we must have 
    \[%
\beginpgfgraphicnamed{nodistlaw1}
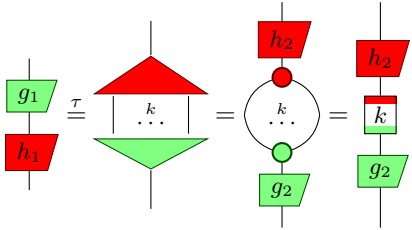
\begin{tikzpicture}
	\begin{pgfonlayer}{nodelayer}
		\node [style=none] (0) at (0, 0.75) {};
		\node [style=green map] (1) at (0, 0.25) {$g_1$};
		\node [style=red map] (2) at (0, -0.5) {$h_1$};
		\node [style=none] (3) at (0, -1) {};
	\end{pgfonlayer}
	\begin{pgfonlayer}{edgelayer}
		\draw (0.center) to (1);
		\draw (2) to (1);
		\draw (2) to (3.center);
	\end{pgfonlayer}
\end{tikzpicture}}
\endpgfgraphicnamed \stackrel{\tau}{=}
\beginpgfgraphicnamed{nodistlaw2}
\InputIfFileExists{nodistlaw2.tikz}{}{\input{./figures/nodistlaw2.tikz}}
\endpgfgraphicnamed = 
\beginpgfgraphicnamed{nodistlaw3}
\InputIfFileExists{nodistlaw3.tikz}{}{\input{./figures/nodistlaw3.tikz}}
\endpgfgraphicnamed = 
\beginpgfgraphicnamed{nodistlaw4}
\InputIfFileExists{nodistlaw4.tikz}{}{\input{./figures/nodistlaw4.tikz}}
\endpgfgraphicnamed
    \]
    for some $g_2\in G$ and $h_2\in H$.  From this we have
    \[
    g_2^\dg h_1g_1h_2^\dg = \mathbf{k}
    \]
    The lefthand side is always unitary, but the righthand side is
    unitary iff $\mathbf{k}$ is invertible in the internal integers
    $R$; hence by an appropriate choice of model---choose
    $R=\mathbb{Z}_k$ for example---this equation does not hold.
  \end{proof}
\end{theorem}

\begin{remark}\label{rem:Euler-decomp}
Note that even if we restrict to our attention to the case where $R$
is a field, then the phase groups provide an obstruction to a distributive law.  
For
example, consider the standard model of
$\IFKd(S^1,\mathbb{Z}_2,S^1,\mathbb{Z}_2)$ over the complex numbers.
In this model the unitaries generate the group $\mathrm{SU}(2)$; if
the distributive law held it would imply that each $u \in
\mathrm{SU}(2)$ could be expressed as two orthogonal rotations, rather
than the known three of Euler decomposition.
\end{remark}






\section{Conclusions and future work}
\label{sec:concl-future-work}

In this paper we have described a sequence of PROPs based on stronger
and stronger interactions between a pair of Frobenius algebras
augmented with phase groups.  At each step another feature of quantum
mechanics is introduced, approaching closer to the full theory.  Since
the each PROP is parameterised by the phase groups, we can view them
as freely constructed quantum-like theories with the given dynamics.
Further, we have shown that, unlike the case of interacting Hopf
algebras \cite{Bonchi2014a}, such theories arise via distributive laws
of PROPs only in special circumstances.

\paragraph{Comparison to ``Interacting Hopf algebras''.}
\label{sec:comparison-bonchi-et}

The similarities between our system and that of Bonchi, Sobocinksi,
and Zanasi \cite{Bonchi2014a} are striking.  Taking the Frobenius
structure as primitive yields \emph{almost} the same theory as
starting with the Hopf structure, and requires fewer axioms to be
imposed.  The main extra ingredient in $\IFA$ are the phase groups,
which play rather badly with the Hopf algebra structure as Lemma
\ref{lem:phases-are-not-bialg} shows; we will ignore them and focus on
$\IFA(1,1)$.
Unlike in \IFA, all the PROPs of \cite{Bonchi2014a} have
trivial scalars; this forces the generating object to be
1-dimensional.

As noted in Section \ref{sec:inter-frob-algebr}, \IFA contains both
\HA and $\HA\op$; however it does not validate any of the axioms
concerning the invertibility of the ring elements, nor their
commutation with the ``wrong'' bialgebra maps\footnote{to wit: (W1),
  (W7), (W8), (B1), (B7), (B8), (S1) and (S2).}.  Here Bonchi et al
rely on the assumption that $R$ is a PID.  However this assumption
fails in, \eg, $\mathbb{CZ}_4$ which is a perfectly good model of
\IFA.  However, in prime dimensional models, Corollary
\ref{cor:primality-gives-unitarity} implies that these axioms are
validated, hence:

\begin{theorem}\label{thm:IH-in-prime-d-IFKd}
Every \emph{\IFKd} algebra of prime dimension includes a
copy of $\mathbb{IH}^w_R$  and this coincides with the image of
$\IFA(1,1)$, modulo scalar factors.
\end{theorem}






\paragraph{Comparison to \zxcalculus.}
\label{sec:comp-zxcalc}

The main inspiration for this approach is the \zxcalculus.  Writing
$S^1$ for the circle group, the PROP
$\IFKd(S^1,\mathbb{Z}_2,S^1 ,\mathbb{Z}_2)$, contains
all the elements and most of the equations of the \zxcalculus, but
there are some key differences.  Firstly, $S^1$ is the entire phase 
group i.e. no new phases are generated by the action of $\mathbb{Z}_2$.
Secondly, the \zxcalculus incorporates the \emph{Hadamard gate}, which
is a definable map which exchanges the colours.  In consequence, the
sets of \gdot- and \rdot-unbiased points are not disjoint in the
\zxcalculus.  We will explore necessary and sufficient conditions for
such a map to exist abstractly in future work; a connection with
Gogioso and Zeng's \cite{Gogioso:2015aa} seems likely.

\paragraph{Further work}
\label{sec:further-work}


Many interesting algebraic properties of \IFA and its relatives remain
unexplored: most notable is role of the semi-direct product in the phase
group (Theorem \ref{thm:semi-direct-product}), and the possibility to
define Hadamard transforms purely abstractly.  A tempting next phase
of development would to investigate topological features, by
considering the case of Lie groups.
Finally we note that in the \emph{models} of the \zxcalculus (although
not derivable) we have the Euler decomposition for $\mathrm{SU}(2)$
giving every unitary as a composition of at most three unitaries.  An
abstract understanding of this would be most valuable.

\small
\bibliography{all}

\newpage
\normalsize


\appendix

\section{Vector Space Models}
\label{sec:vector-space-models}

A source of examples of Frobenius algebras are group algebras which for the 
purposes of this paper means the following:

\begin{definition}
Let $K$ be a field and $G$ a finite group such that the characteristic of 
$\mathbbm{k}$ 
does not divide $|G|$. The \emph{group algebra} is defined to be a 
$\mathbbm{k}$-vector space $\mathbbm{k}G$ with a basis elements $\ket{g}$ for 
each $g \in G$ 
equipped with the following data: 

\begin{itemize}
 \item an algebra structure $\rmu : \mathbbm{k}G \otimes \mathbbm{k}G 
\to \mathbbm{k}G$ defined point-wise on the basis elements by the group 
mulitpication, 
i.e. $\ket{g} \otimes \ket{h} \mapsto \ket{gh}$ and with unit $\reta: K \to 
\mathbbm{k}G$ given by the identity element in $G$, $1 \mapsto \ket{e}$.

\item a coalgebra structure $\rdelta : \mathbbm{k}G \to \mathbbm{k}G \otimes 
\mathbbm{k}G$ given by $\ket{g} 
\mapsto \frac{1}{|G|} \sum\limits_{hh' = g} \ket{h} \otimes \ket{h'}$ with 
counit $\repsilon: \mathbbm{k}G \to \mathbbm{k}$ given by $\ket{g} \mapsto 
\left\{
  \begin{array}{lr}
    1 & \text{if } g = e\\
    0 & \text{if } g \not= e
  \end{array}
\right.
$

\item an algebra structure $\gmu : \mathbbm{k}G \otimes \mathbbm{k}G \to 
\mathbbm{k}G$ given by $\ket{g} 
\otimes \ket{h} \mapsto \left\{
  \begin{array}{lr}
    \ket{g} & \text{if } g=h\\
    0 & \text{if } g \not=h 
  \end{array}
\right.$ with unit $\geta : \mathbbm{k} \to \mathbbm{k}G$ given by $1 \mapsto 
\sum\limits_{g \in 
G}\ket{g}$

\item a coalgebra structure given by $\gdelta : \mathbbm{k}G \to \mathbbm{k}G 
\otimes \mathbbm{k}G$ given by 
$\ket{g} \mapsto \ket{g} \otimes \ket{g}$ with counit $\gepsilon : \mathbbm{k}G 
\to \mathbbm{k}$ 
given by $\ket{g} \mapsto 1$ for all $g \in G$.
\end{itemize}
\end{definition}

Group algebras come equipped with a canonical hopf
algebra structure $(\rmu, \gdelta)$, often called the \emph{convolution 
algebra}. From
the following theorem of Maschke, group algebras often carry a
canonical special frobenius algebra structure also.

\begin{theorem}[Maschke]
  Let $G$ be a finite group with $|G|=D$, and $\mathbbm{k}$ a field. If the
  characteristic of $\mathbbm{k}$ does not divide $D$ then the group algebra
  $\mathbbm{k}G$ is a special frobenius algebra with comultiplication and counit

\begin{equation*}
\rdelta::\ket{g} \mapsto \frac{1}{|G|}\big(\sum\limits_{hh'=g}\ket{h} \otimes 
\ket{h'}\big) , \ \ \ \ \ \ \ \ \repsilon :: g \mapsto 
  \begin{cases} 
   D & \text{if } g = e \\
   0 & \text{otherwise}
  \end{cases}
\end{equation*}

\end{theorem}

The prototypical examples of $\dg$-PROPs of interacting frobenius
algebras arise from finite groups and group algebras. Let $G$ be a
finite abelian group with $|G|=D$, and $\mathbbm{k}$ a field such that
$\text{char}(\mathbbm{k}) \nmid |G|$. Letting $D= \overbrace{1_K+...+1_K}^{D}$,
then if $\sqrt{D}$ exists in $K$ then we obtain a $\dg$-PROP of
interacting frobenius algebras generated by the underlying vector
space $\mathbbm{k}G$, with monoidal product the tensor product of vector
spaces. Morphisms are generated by the following
mulitplication/comultiplication maps along with their respective
units/counits, as well as the antipode $s:: \ket{g} \mapsto \ket{
  g^{-1}}$.

Given that the vector space has a chosen basis $\{\ket{g} \}_{g \in
  G}$ we define the dagger via the compact structure i.e. red cups and
caps. For any linear map $f:\mathbbm{k}G^{\otimes m} \to \mathbbm{k}G^{\otimes 
n}$ we
define $f^{\dg} = f\gtrans$. It is easy to verify that $\rmu^{\dg} =
\rdelta$, $\gmu^{\dg} = \gdelta$.

\begin{remark}
  It might not be the case the dimension $D$ has a square root in the
  underlying field. Consider $G= \mathbb{Z}_2$ with $\mathbbm{k}$ the finite
  field $\mathbb{F}_5$. As long as the characteristic of $\mathbbm{k}$ does not
  divide $G$ we will have a pair of special commutative frobenius
  algebras $(\gmu,\gdelta)$ as above, and by defining $\rmu::
  \ket{g}\otimes \ket{h} \mapsto \ket{gh}$ and $\rdelta :: \ket{g}
  \mapsto \frac{1}{D} \big( \sum\limits_{hh'=g} \ket{h} \otimes
  \ket{h'} \big)$. These are also jointly hopf algebras but the
  $\dg$-condition does not hold on the nose, but up to a scalar
  factor.
\end{remark}

The following definition generalises the \emph{dual group} of Pontryagin 
duality.

\begin{definition}
  Let $G$ be a group, $\mathbbm{k}$ a field, and let $\mathbbm{k}^{\times}$ 
denote the
  multiplicative group of non-zero elements in $\mathbbm{k}$. The
  \emph{$\mathbbm{k}$-characters} of $G$ are the group homomorphisms $G \to
  \mathbbm{k}^{\times}$. The collection of $\mathbbm{k}$-characters forms a 
group under
  pointwise addition in $\mathbbm{k}$ which we will call the 
\emph{$\mathbbm{k}$-dual
    group} $G^{\wedge \mathbbm{k}}$. We may refer to $\mathbbm{k}$-characters 
simply as
  \emph{characters} if there is no risk of ambiguity.
\end{definition}

The $\mathbb{C}$-dual group of $G$ is then the dual group in the usual sense of 
Pontryagin duality.

\begin{lemma}\label{lem:characters-are-set-like-for-group-algs}
  For a group algebra $\mathbbm{k}G$ the set-like elements of $\rdelta$ are in
  one-to-one corresondence with $\mathbbm{k}$-characters. In particular, the
  set-like elements are of the form $\sum\limits_{g \in G}
  \chi(g)\ket{g}$ for each group character $\chi: G \to \mathbbm{k}^{\times}$.
\begin{proof}
  Let $\chi: G \rightarrow \mathbbm{k}^{\times}$ be a character, the claim is
  that the element $p = \sum\limits_{g \in G} \chi(g)\ket{g}$ is a
  set-like element for $\rdelta$.
\begin{align*}
\rdelta(p) &= \sum\limits_{g \in G} \chi(g)\rdelta(\ket{g}) \\
    &= \sum\limits_{g \in G} \chi(g) \sum\limits_{hh'=g} \ket{h} \otimes 
\ket{h'} \\
    &= \sum\limits_{h,h' \in G} \chi(hh')|h\rangle \otimes \ket{h'} \\
    &= \sum\limits_{h,h' \in G} \chi(h)\chi (h')|h\rangle \otimes \ket{h'} \\
    &= p \otimes p
\end{align*}
as required.

Conversely suppose $p = \sum\limits_{g \in G} \alpha_g \ket{g}$ is a
set-like element. We have

\begin{equation*}
p \otimes p = \sum\limits_{h,h' \in G} \alpha_h \alpha_{h'} \ket{h} \otimes 
\ket{h'}
\end{equation*}
and we have
\begin{align*}
\rdelta(p) &= \sum\limits_{g \in G} \alpha_g \delta( \ket{g}) \\
    &= \sum\limits_{g \in G} \alpha_g \sum\limits_{hh'=g} \ket{h} \otimes 
\ket{h'} \\
    &= \sum\limits_{h,h' \in G} \alpha_{hh'}\ket{h} \otimes \ket{h'}
\end{align*}
by assumption that $p \otimes p = \gdelta(p)$ we conclude that
$\alpha_h\alpha_{h'} = \alpha_{hh'}$ for all $h$ and $h'$ in $G$. In
particular $\alpha_h\alpha_e = \alpha_h$ for all $h$ and hence
$\alpha_e = 1$. Hence the map $g \mapsto \alpha_g$ is a group
homomorphism, as required.
\end{proof}

\end{lemma}

\begin{corollary}
  There are fields $\mathbbm{k}$ such that there are $\dg$-SCFAs in
  $\emph{\textbf{Vect}}_{\mathbbm{k}}$ which do not have enough set-like 
elements.

\begin{proof}
  Consider the group $\mathbb{Z}_4$ and the field $\mathbb{R}$. Since
  $\text{char}(\mathbb{R})=0$ the group algebra $(\rmu,\rdelta)$ is a
  special commutative frobenius algebra. Now the claim is that there
  are only two group homomorphisms $\mathbb{Z}_4 \to
  \mathbb{R}^{\times}$. Let $\chi$ be such a homomorphism then
  $\text{im}(\chi)$ is a finite subgroup of $\mathbb{R}^{\times}$, but
  there is only one non-trivial finite subgroup of $\mathbb{R}$ namely
  $\{ \ 1, -1 \ \} \iso \mathbb{Z}_2$ hence we can characterise all of
  the group characters as the homomorphisms $\mathbb{Z}_4 \to
  \mathbb{Z}_2$ of which there are only two.

  Hence the comultiplication $\rdelta$ does not have enough set-like
  elements.
\end{proof}
\end{corollary}

\begin{definition}
  The \emph{exponent} $\emph{exp}(G)$ of a finite group $G$ is defined
  to be $\emph{max}_{g \in G}\{ \ \emph{ord}(g) \ \}$ where the
  \emph{order of an element} $\emph{ord}(g)$ is defined to be the
  smallest positive integer $n$ such that $g^n = e$.
\end{definition}

\begin{lemma}
The order of each element in $G$ divides the exponent of $G$.
\end{lemma}

\begin{theorem}
  Let $G$ be a group with $\emph{exp}(G)=d$, and let $\mathbbm{k}$ be a field
  such that \emph{char}$(\mathbbm{k})$ does not divide $|G|$. For the group
  algebra $\mathbbm{k}G$ the following are equivalent:
\begin{itemize}
 \item $\rdelta$ has enough set-like elements.
 \item $\mathbbm{k}$ is a splitting field for $x^d - 1$.
\end{itemize}
\begin{proof}
  It is easy to see that $x^d-1$ splits in $\mathbbm{k}$ iff $\mathbbm{k}$ has 
$d$
  distinct $d^{\text{th}}$ roots of unity. There are at most $d$
  $d^{\text{th}}$-roots of unity and hence they form a finite subgroup
  of $\mathbbm{k}^{\times}$. Finite subgroups of $\mathbbm{k}^{\times}$ are 
cyclic for all
  fields, hence $x^d-1$ splitting in $\mathbbm{k}$ is equivalent to 
$\mathbbm{k}$ having a
  subgroup isomorphic to $\mathbb{Z}_d$.

  Suppose $\mathbb{Z}_d \leq K^{\times}$, then for any group
  homomorphism $\varphi: G \to \mathbbm{k}^{\times}$, the image of $\phi$ must
  be a subgroup of $\mathbb{Z}_d$. Hence it is enough to characterise
  all of the group homomorphisms $G \to \mathbb{Z}_d$.

  Every finite abelian group can be written uniquely in the form
  $\mathbb{Z}_{n_1} \oplus ... \oplus \mathbb{Z}_{n_k}$ where each
  $n_i$ divides $n_{i+1}$, known as the invariant factor
  decompositon. Clearly we have $n_k = d$. By the universal property
  of coproducts every homomorphism $G\to \mathbb{Z}_d$ is determined
  by a tuple of group homomorphisms $(\varphi_{n_i}: \mathbb{Z}_{n_i}
  \to \mathbb{Z}_d)_{1 \leq i \leq k}$.

  If $n$ divides $d$ then there are exactly $n$ group homomorphisms
  $\mathbb{Z}_n \to \mathbb{Z}_d$. Hence there are
  $\prod\limits_{1\leq i \leq k} n_{i}= |G|$ homomorphisms $G \to
  \mathbb{Z}_d$. Hence if $x^d-1$ splits there are $|G|$ homomorphisms
  $G\to K^{\times}$.

  Conversely, suppose that $K$ has fewer than $d$
  $d^{\text{th}}$-roots of unity i.e. that $\mathbbm{k}^{\times}$ does not
  contain a subgroup isomorphic to $\mathbb{Z}_d$. The image of any
  homomorphism $\mathbbm{k}^{\times}$ must be a cyclic group of order $m$ such
  that $m$ divides $d$. For the collection of homomorphisms
  $(\varphi_j: G \rightarrow \mathbbm{k}^{\times})_{1\leq j \leq l}$ for some
  $l$. let $m' = \emph{max}_{1\leq j \leq l}\{ \
  |\emph{im}(\varphi_{j})| \ \}$. Then the image of every $\varphi_j$
  lies within the subgroup isomorphic to $\mathbb{Z}_m$ and hence the
  group homomorphisms $G \to \mathbbm{k}^{\times}$ are completely characterised
  by the group homomorphisms $G \to \mathbb{Z}_{m'}$. Since
  $\mathbb{Z}_d$ appears as a direct summand in the invariant factor
  decompositon of $G$ and by assumption $m' < d$, there are strictly
  less than $\prod\limits_{1\leq i \leq k} n_{i}= |G|$ homomorphisms
  $G \to \mathbb{Z}_{m'}$. Required result is the contrapositive
  statement.
\end{proof}
\end{theorem}



\section{Proofs}
\label{sec:proofs}



{
\def\thetheorem{\ref{thm:hopf-iff-units-are-real}}
\begin{theorem}
  The morphisms $(\gdelta,\gepsilon,\rmu,\reta)$ form a Hopf algebra if and only
  if $\reta = (\repsilon)\gtrans$ and $\gepsilon = (\geta)\rtrans$, \ie
  \begin{equation}
    \label{eq:BAplusA}\tag{+}
\beginpgfgraphicnamed{gunit-rreal}
\InputIfFileExists{gunit-rreal.tikz}{}{\input{./figures/gunit-rreal.tikz}}
\endpgfgraphicnamed = \gcounit
    \qquad\qquad
\beginpgfgraphicnamed{runit-greal}
\InputIfFileExists{runit-greal.tikz}{}{\input{./figures/runit-greal.tikz}}
\endpgfgraphicnamed = \runit
  \end{equation}
\begin{proof}
\[
\beginpgfgraphicnamed{hopf-lawb}
\InputIfFileExists{hopf-lawb.tikz}{}{\input{./figures/hopf-lawb.tikz}}
\endpgfgraphicnamed \quad = \quad %
\beginpgfgraphicnamed{antipode-equiv1}
\InputIfFileExists{antipode-equiv1.tikz}{}{\input{./figures/antipode-equiv1.tikz}}
\endpgfgraphicnamed \quad = \quad 
\beginpgfgraphicnamed{antipode-equiv2}
\InputIfFileExists{antipode-equiv2.tikz}{}{\input{./figures/antipode-equiv2.tikz}}
\endpgfgraphicnamed \quad = \quad %
\beginpgfgraphicnamed{antipode-equiv3}
\InputIfFileExists{antipode-equiv3.tikz}{}{\input{./figures/antipode-equiv3.tikz}}
\endpgfgraphicnamed \quad = \quad 
\beginpgfgraphicnamed{antipode-equiv4}
\InputIfFileExists{antipode-equiv4.tikz}{}{\input{./figures/antipode-equiv4.tikz}}
\endpgfgraphicnamed
\]
\end{proof}
\end{theorem}
}

{
\def\thetheorem{\ref{cor:antipode-and-real-points}}
\begin{corollary}
  Suppose $k:0\to 1$ is \gdot-real, and let $h=\Lambda\rconj (k)$.  Then
  \begin{enumerate}
  \item $s\circ k = k\rconj$, and
  \item $s\circ h \circ s = h^\dg$.
  \end{enumerate}
  \begin{proof}
1. Since $k$ is \gdot-real we have $s\circ k = k\grtrans =
      {k^\dg}{}\rtrans = k\rconj$.
      
      2. Since $s$ commutes with $\rmu$,
      we have $s\circ \Lambda\rconj (k) \circ s = \Lambda\rconj(s\circ
      k) = \Lambda\rconj(k\rconj)$ which gives the result by Lemma
      \ref{lem:phases-and-unb-points}.
  \end{proof}
\end{corollary}
}

{
\def\thetheorem{\ref{prop:internal-ints}}
\begin{proposition}
  Let $f$ be a bialgebra morphism, and $s$ the antipode; then for all
  $g,h:A\to A$ we have:
  \begin{enumerate}
  \item $f \circ (g + h) = (f\circ g) + (f\circ h)\;$,
  \item $(g + h) \circ f = (g\circ f) + (h\circ f)\;$, and
  \item  $f + (f\circ s) = 0\;$. 
  \end{enumerate}
\begin{proof} 1.
    \[
     f \circ (g + h) \quad = \quad %
\beginpgfgraphicnamed{BialgRingElmts1}
\InputIfFileExists{BialgRingElmts1.tikz}{}{\input{./figures/BialgRingElmts1.tikz}}
\endpgfgraphicnamed \quad = \quad 
\beginpgfgraphicnamed{BialgRingElmts2}
\InputIfFileExists{BialgRingElmts2.tikz}{}{\input{./figures/BialgRingElmts2.tikz}}
\endpgfgraphicnamed \quad = \quad (f\circ g) + (f\circ h)
    \]
    2.
\[
     (g + h) \circ f \quad  =\quad %
\beginpgfgraphicnamed{BialgRingElmts4}
\InputIfFileExists{BialgRingElmts4.tikz}{}{\input{./figures/BialgRingElmts4.tikz}}
\endpgfgraphicnamed \quad =  \quad 
\beginpgfgraphicnamed{BialgRingElmts3}
\InputIfFileExists{BialgRingElmts3.tikz}{}{\input{./figures/BialgRingElmts3.tikz}}
\endpgfgraphicnamed \quad = \quad (g\circ f) + (h\circ f)
    \]
    3.
\[
f + (f\circ s)\quad = \quad %
\beginpgfgraphicnamed{BialgRingElmts5}
\InputIfFileExists{BialgRingElmts5.tikz}{}{\input{./figures/BialgRingElmts5.tikz}}
\endpgfgraphicnamed \quad = \quad 
\beginpgfgraphicnamed{BialgRingElmts6}
\InputIfFileExists{BialgRingElmts6.tikz}{}{\input{./figures/BialgRingElmts6.tikz}}
\endpgfgraphicnamed \quad = \quad %
\beginpgfgraphicnamed{BialgRingElmts7}
\InputIfFileExists{BialgRingElmts7.tikz}{}{\input{./figures/BialgRingElmts7.tikz}}
\endpgfgraphicnamed \quad = \quad 
\beginpgfgraphicnamed{BialgRingElmts8}
\begin{tikzpicture}
	\begin{pgfonlayer}{nodelayer}
		\node [style=red vertex] (0) at (0, -0) {};
		\node [style=green vertex] (1) at (0, 0.5) {};
		\node [style=none] (2) at (0, 1) {};
		\node [style=none] (3) at (0, -0.5) {};
	\end{pgfonlayer}
	\begin{pgfonlayer}{edgelayer}
		\draw [style=simple] (1) to (2.center);
		\draw [style=simple] (0) to (3.center);
	\end{pgfonlayer}
\end{tikzpicture}}
\endpgfgraphicnamed \quad = \quad 0
\]
  \end{proof}
\end{proposition}
}
{
\def\thetheorem{\ref{lem:inv-implies-bialg}}
\begin{lemma}
If $f$ is an invertible bialgebra morphism then $f^{-1}$ is also a bialgebra 
morphism.

\begin{proof}
Suppose $f$ is a bialgebra morphism and consider $ f^{-1} \circ \mu = f^{-1} 
\circ \mu \circ (f  \otimes f )\circ (f^{-1} \otimes f^{-1}) = f^{-1} \circ f 
\circ \mu \circ (f^{-1} \otimes f^{-1})= \mu \circ (f^{-1} \otimes f^{-1})$, as 
required.
\end{proof}
\end{lemma}
}

{
\def\thetheorem{\ref{lem:classical-commutes-with-structure}}
\begin{lemma}
  If $h:1\to 1$ is \gdot-classical then it commutes with
  $\gdelta$ and $\gmu$.
  \begin{proof}
    Let $k$ be the \gdot-classical point such that
    $h=\Lambda\rconj(k)$.  Then
    \[
\beginpgfgraphicnamed{clarefrobhom4}
\InputIfFileExists{clarefrobhom4.tikz}{}{\input{./figures/clarefrobhom4.tikz}}
\endpgfgraphicnamed \ \ = \ \ %
\beginpgfgraphicnamed{clarefrobhom3}
\InputIfFileExists{clarefrobhom3.tikz}{}{\input{./figures/clarefrobhom3.tikz}}
\endpgfgraphicnamed \ \ = \
    \ %
\beginpgfgraphicnamed{clarefrobhom1}
\InputIfFileExists{clarefrobhom1.tikz}{}{\input{./figures/clarefrobhom1.tikz}}
\endpgfgraphicnamed \ \ = \ \ %
\beginpgfgraphicnamed{clarefrobhom2}
\InputIfFileExists{clarefrobhom2.tikz}{}{\input{./figures/clarefrobhom2.tikz}}
\endpgfgraphicnamed \ \ =
    \ \ %
\beginpgfgraphicnamed{clarefrobhom5}
\InputIfFileExists{clarefrobhom5.tikz}{}{\input{./figures/clarefrobhom5.tikz}}
\endpgfgraphicnamed
    \]
hence $h$ commutes with \gdelta, and dually $h^\dg$ commutes with
$\gmu$. The result follows by Corollary
\ref{cor:antipode-and-real-points}.2.
\end{proof}
\end{lemma}
}

\section{Iterated Distributive Laws}
\label{sec:IteratedDistLaws}
Given monads $T$ and $S$, the existence of a 
distributive law $\lambda: T  S \to S  T$ is a sufficient 
condition which ensures that the composite $S  T$ is again a monad. If instead 
we are given three monads $A$, $B$, and $C$ with distributive laws 
\[ 
\lambda_1 : BA \to AB
\]
\[
\lambda_2 : CA \to AC 
\]
\[
 \lambda_3: CB \to BC
\]
it is natural to ask the question: under what conditions is the composition 
$ABC$ a monad? 

\begin{definition}\label{def:DistSeriesOfMonads}
 A \emph{distributive series of monads} is a family of monads $\{T_i\}_{1\leq 
i \leq n}$ for some $n \geq 3$ with 
distributive laws $\lambda_{ij}: T_i T_j \to T_j T_i$ for each 
$j > i$, such that 
for all $i>j>k$ the Yang-Baxter diagram commutes.
\[
\begin{tikzpicture}
\node(A) at (0,0) {$T_i T_j T_k$};
\node(B) at (2.5,1) {$T_j T_i T_k$};
\draw[->](A) to node [above]{$\lambda_{ij}T_k$}(B);
\node(C) at (5,1) {$T_j T_k T_i$};
\draw[->](B) to node [above]{$T_j\lambda_{ik}$}(C);
\node(D) at (7.5,0) {$T_k T_j T_i $};
\draw[->](C) to node [above]{$\lambda_{jk}T_i$}(D);
\node(E) at (2.5,-1) {$T_i T_k T_j$};
\draw(A)[->] to node [below]{$T_i \lambda_{jk}$}(E);
\node(F) at (5,-1) {$T_k T_i T_j$};
\draw[->](E) to  node [below]{$\lambda_{ik} T_j$}(F);
\draw[->](F) to  node [below]{$T_k \lambda_{ji} $}(D);
\end{tikzpicture}
\]
\end{definition}

\begin{theorem}[Cheng]
Fix $n \geq 3$. For a distributive series of monads $\{T_i\}_{1\leq i \leq n}$ 
then for each $1\leq j <n $
the composition $T_1 T_2... T_j$ is a monad, and the compostition $T_{j+1} 
T_{j+2} ... T_{n}$ is a monad.
\begin{proof}
\cite[Theorem 2.1]{Cheng2011PSP8239257}
\end{proof}
\end{theorem}

Then in the above example with monads $A,B$ and $C$, if the distributive laws 
$\lambda_1$, $\lambda_2$ and $\lambda_3$ satisfy the Yang-Baxter diagram then 
the composition $ABC$ is also a monad.

\end{document}